\pdfoutput=1  

\documentclass[aps,superscriptaddress,floatfix,nofootinbib,preprintnumbers,eqsecnum,twocolumn]{revtex4-1}
\usepackage{lipsum}
\usepackage{graphicx}
\usepackage{amssymb,amsmath,multirow}
\usepackage{comment,color,placeins,amsmath,amsfonts,enumitem}

\usepackage{hyperref}
\usepackage{xcolor}
\hypersetup
{
  colorlinks,
  linkcolor={blue!80!black},
  citecolor={blue!70!black},
  urlcolor={blue!70!black}
}

\makeatletter
\renewcommand*\env@matrix[1][\arraystretch]{%
  \edef\arraystretch{#1}%
  \hskip -\arraycolsep
  \let\@ifnextchar\new@ifnextchar
  \array{*\c@MaxMatrixCols c}}
\makeatother

\newcommand{\biggg}{\bBigg@{3}}
\newcommand{\Biggg}{\bBigg@{4}}
\makeatother

\def\beq{\begin{equation}}
\def\eeq{\end{equation}}
\def\beqn{\begin{eqnarray}}
\def\eeqn{\end{eqnarray}}
\def\calM{{\cal M}}

\def\half{{\textstyle{1\over 2}}}

\def\ie{{\it i.e.}\/}
\def\eg{{\it e.g.}\/}

\def\mbar{{\overline{m}}}

\def\rhobar{{\overline{\rho}}}
\def\fd{{\rm 4D}}

\def\lambdabar{{\overline{\lambda}}}

\def\etabar{{\overline{\eta}}}

\def\fhX{{\hat{f}_X}}

\def\IR{\relax{\rm I\kern-.18em R}}
 \font\cmss=cmss10 \font\cmsss=cmss10 at 7pt
\def\IQ{\relax{\rm I\kern-.18em Q}}
\def\IZ{\relax\ifmmode\mathchoice
 {\hbox{\cmss Z\kern-.4em Z}}{\hbox{\cmss Z\kern-.4em Z}}
 {\lower.9pt\hbox{\cmsss Z\kern-.4em Z}}
 {\lower1.2pt\hbox{\cmsss Z\kern-.4em Z}}\else{\cmss Z\kern-.4em Z}\fi}

\newcommand{\alignedfrac}[2]{%
    \setbox0\hbox{$#1$}        
    \dimen0=\wd0               
    \setbox1\hbox{$#2$}        
    \dimen1=\wd1               
    \ifdim\wd0<\wd1            
        \dfrac{#1\hfill}{#2}   
    \else                      
        \dfrac{#1}{#2\hfill}   
    \fi
}

\newcommand{\p}{\partial}
\newcommand{\ve}[1]{{\mathbf{#1}}}
\newcommand{\nn}{\nonumber}
\newcommand{\abs}[1]{\left| #1 \right|}

\newcommand{\expt}[1]{\left\langle #1 \right\rangle}

\newcommand{\be}{\begin{equation}}
\newcommand{\ee}{\end{equation}}
\newcommand{\ba}{\begin{align}}
\newcommand{\ea}{\end{align}}

\newcommand{\il}[1]{\mbox{$#1$}}

\def\ie{{\it i.e.}\/}
\def\eg{{\it e.g.}\/}

\newcommand{\newc}{\newcommand}
\newc{\gsim}{\lower.7ex\hbox{$\;\stackrel{\textstyle>}{\sim}\;$}}
\newc{\lsim}{\lower.7ex\hbox{$\;\stackrel{\textstyle<}{\sim}\;$}}

\begin{document}
\title{Kaluza-Klein Towers in the Early Universe: \\ Phase Transitions, Relic Abundances, and Applications to Axion Cosmology} 
\author{Keith R. Dienes}
\email{dienes@email.arizona.edu}
\affiliation{Department of Physics, University of Arizona, Tucson, AZ 85721 USA}
\affiliation{Department of Physics, University of Maryland, College Park, MD 20742 USA}
\author{Jeff Kost}
\email{jkost@email.arizona.edu}
\affiliation{Department of Physics, University of Arizona, Tucson, AZ 85721 USA}
\author{Brooks Thomas}
\email{thomasbd@lafayette.edu}
\affiliation{Department of Physics, Lafayette College, Easton, PA 18042 USA}

\begin{abstract}
\noindent
We study the early-universe cosmology of a Kaluza-Klein (KK) tower of scalar fields 
in the presence of a mass-generating phase transition, focusing on the time-development 
of the total tower energy density (or relic abundance) as well as its distribution across 
the different KK modes.  We find that both of these features are extremely sensitive 
to the details of the phase transition and can behave in a variety of ways significant 
for late-time cosmology.  In particular, we find that the interplay between the temporal 
properties of the phase transition and the mixing it generates are responsible for both 
enhancements and suppressions in the late-time abundances, sometimes by many orders
of magnitude.  We map out the complete model parameter space and determine where 
traditional analytical approximations are valid and where they fail.  In the latter 
cases we also provide new analytical approximations which successfully model our results.
Finally, we apply this machinery to the example of an axion-like field in the bulk, 
mapping these phenomena over an enlarged axion parameter space that extends beyond that
accessible to standard treatments.  An important by-product of our analysis is the 
development of an alternate ``UV-based'' effective truncation of KK theories which has 
a number of interesting theoretical properties that distinguish it from the more 
traditional ``IR-based'' truncation 
typically used in the extra-dimension literature.
\end{abstract}

\maketitle




\FloatBarrier
\section{Introduction\label{sec:Intro}}


The presence of additional light scalar degrees of freedom is a common feature of many 
extensions of the Standard Model (SM).  Examples include the 
QCD axion~\cite{PecceiQuinn1,PecceiQuinn2,WeinbergAxion,WilczekAxion}, a more
general set of axions and axion-like particles~\cite{LowEnergyFrontier}, 
majorons~\cite{ChikashigeMajorons,GelminiMajorons}, familons~\cite{WilczekFamilon}, 
chameleons~\cite{RatraChameleons,KhouryChameleons1,KhouryChameleons2,BraxChameleons},
branons~\cite{DobadoBranon}, dilatons, and a variety of other string and geometric 
moduli~\cite{Polchinski,BBS,GSW}.  Scalars of this sort are typically 
light due to symmetries of the high-scale theory which prevent them from acquiring masses.
As a result, masses for these particles must be generated by dynamical processes at lower scales 
which break these symmetries --- often in conjunction with a cosmological
phase transition.  This dynamics also generically leads to a non-trivial time-dependence for 
the masses of these fields, the details of which can have a significant impact on  
their late-time energy densities.  Understanding these effects is crucial, since    
such particles can potentially have a variety of phenomenological and cosmological 
consequences, both helpful and harmful.  Indeed, scalars of this sort can 
contribute to the total abundances of dark matter or dark energy; induce additional, 
late periods of reheating after cosmic inflation, potentially resulting in substantially 
modified cosmologies~\cite{WatsonNonThermalHistories};  
prematurely induce matter-domination and/or overclose the universe; 
and disrupt the formation of light elements through late decays. 

While the detailed dynamics of mass generation can have important consequences
even in scenarios involving only a single light scalar, the situation becomes far 
richer in scenarios involving multiple such fields.  
Indeed, in cases in which these scalars have the same quantum numbers, the 
dynamics of mass generation can lead them to experience time-dependent mixing.
This mixing in turn leads to a 
continual redistribution of energy density among the scalars during the mass-generation
epoch.  Such a redistribution can also have profound effects on the late-time energy densities
of these fields, several of which were pointed out in Ref.~\cite{TwoTimescales}.  Specifically,
even in a theory involving only two scalars, such a redistribution can lead their late time 
abundances to experience either an enhancement or a suppression,  depending on
the choice of model parameters.  Moreover, in certain regimes, the system may experience
a parametric resonance which can have further dramatic effects on these abundances.
Finally, the system may also experience a ``re-overdamping'' phenomenon 
in which its energy density reverts to behaving like vacuum energy 
at a time after its initial transition from the overdamped to underdamped regime.

A variety of scenarios for new physics predict large numbers of scalar fields with 
similar quantum numbers.  Indeed, such collections of fields emerge naturally from a 
variety of string constructions~\cite{WittenStringAxion,SvrcekWitten}, from axiverse 
considerations~\cite{Axiverse}, and more 
generally from theories with extra dimensions.  Of course, many features of such 
theories are highly model-dependent, including the number of fields present, 
the form of the mass-squared matrix for these fields (both at very early and at very 
late times), and the time or temperature scales associated with the dynamical mechanisms 
for mass generation.  However,
a given specific 
new-physics context is likely to lead to certain theoretical structures in which
some of these features become fixed.
For example, theories in which a scalar propagates in extra spacetime dimensions  give 
rise to infinite numbers of scalar fields which are organized into Kaluza-Klein (KK) 
towers in which the form of the mass matrix at early times --- even prior to the onset 
of any mass-generating phase transition --- is dictated by the geometry of the 
compactification manifold.  Because such KK towers arise generically in scenarios 
involving compact extra dimensions (such as string theories, supergravity theories, and 
many other high-scale unification models),
the features associated with the cosmological evolution of such KK towers --- including 
the dynamical flow of energy amongst the different KK modes prior to, during, and after
cosmological phase transitions --- are likely to play an important role in 
early-universe cosmology.

In this paper we tackle this important question.  In particular, we study
the early-universe cosmology of a KK tower of scalar fields in the presence
of a mass-generating phase transition, focusing on the time-development of the total 
tower energy density (which determines the relic abundance) as well as its distribution across the different 
KK modes.  On the one hand, this study is critical for understanding the cosmological 
issues associated with extra spacetime dimensions {\it per se}\/.  On the other hand, 
this study also represents an extension of the analysis presented in 
Ref.~\cite{TwoTimescales}.  In that analysis, only two scalar fields were considered, 
but these fields were permitted to experience the most general possible
mixings between them.  In this analysis, by contrast, the number of scalar fields is 
infinite but the structure of the KK tower and the interplay between brane and bulk 
physics imposes a strict mixing structure.  A goal of this paper is therefore to 
understand the consequences of this simultaneous extension in the number of
fields and narrowing of the allowed mixing structure. 

As we shall see, studying the dynamics of a complete KK tower can be a significant
undertaking due to its infinite number of fields --- especially in the presence of 
a non-trivial, time-dependent, mass-generating and mixing-inducing phase transition.
Therefore, in this paper, we shall work towards this goal in distinct stages by
starting with only one KK mode and then introducing more and more KK excitations into 
the system.  Ultimately, we shall build towards the full, infinite  KK tower.
In each case, our goal is to trace the flow of energy through the system and to examine 
the behavior of such quantities as the total late-time energy density as well as its 
distribution across the available KK modes.  Moreover, our goal is to retain as much 
generality as possible.  We shall therefore refrain from specifying the exact nature 
of our KK tower other than to specify that its constituents are scalars.   Our results 
will therefore be broadly applicable to any of the particular scalar fields listed above.

This paper is organized as follows.  In Sect.~\ref{sec:TheModel}, we begin by establishing 
the context in which we shall be working and review the general properties of our setup.
Then, in Sect.~\ref{sec:FourDimensionalLimit}, we begin our study by considering the 
case of only a single scalar field undergoing a mass-generating phase transition.
In particular,
our analysis will extend beyond the usual ``adiabatic'' and ``abrupt'' regimes commonly 
discussed in the literature.
In Sect.~\ref{sec:FiniteModeNumber}, we then broaden our perspective by
considering the case of only the $N$ lowest-lying KK excitations, where $N$
is arbitrary but finite.  As we shall see, situations with finite $N$ have a number of 
properties which render their cosmological evolution somewhat distinct.
In Sect.~\ref{Approaching}, we then examine how our system behaves as a function of 
$N$ for \il{N\gg 1}, 
and in Sect.~\ref{sec:KKLimit} we present our results for the full, infinite KK tower.
In order to distinguish this case from the large-$N$ case
in Sect.~\ref{Approaching}, we shall refer to the KK tower as representing the case with \il{N=\infty}.
Until this point, our analysis is completely general and applicable to any scalar
field.  However, in Sect.~\ref{sec:AxionintheBulk}, we then apply our machinery to the case
in which our scalar is an axion or axion-like particle.  This allows us to interpret our 
results within a particular phenomenologically relevant context, and illustrate the 
significant phenomenological implications that our results can have.  Finally, in 
Sect.~\ref{sec:Conclusions}, we summarize our main conclusions and discuss further 
potential consequences of our results.

As evident from this outline, our method for studying the properties of a full KK theory 
involves truncating the KK tower to its $N$ lowest-lying modes and then considering the 
\il{N\to\infty} limit.  Indeed, this is a valid, standard approach.  However, as a by-product 
of our analysis, we shall discover that such a truncation is not unique.  Specifically, we 
shall develop an alternate, ``UV-based'' effective truncation of KK theories which has a 
number of interesting theoretical properties that distinguish it from the more traditional 
``IR-based'' truncation 
typically used in the extra-dimension literature.  Both truncations lead to the same 
physics as \il{N\to\infty}, but differ significantly for finite $N$.
This alternate truncation will be discussed in Sect.~\ref{alttrunc}.

\FloatBarrier
\section{The Framework\label{sec:TheModel}}


In this section, we establish the framework in which our analysis
will take place.
This framework is similar to that considered in Refs.~\cite{DDGAxions,DDM1,DDM2,DDMAxion}, 
and is essentially a KK-oriented extension
of the framework considered in Ref.~\cite{TwoTimescales}. 
Specifically, we consider a flat, five-dimensional spacetime geometry
of the form \il{{\cal M} \times S^1/\IZ_2}, where
${\cal M}$ denotes our usual four-dimensional Minkowski spacetime and
where $S^1/\IZ_2$ denotes an orbifolded circle (\ie, a line segment) of 
length $2\pi R$.
Our spacetime coordinates shall be denoted \il{(x^\mu,x^5)}, \il{\mu=0,1,2,3}, 
where \il{0\leq x^5 \leq 2\pi R} and where the orbifold action is given by \il{x^5\to -x^5}.
We shall further imagine that
our usual Standard-Model (SM) fields and interactions are confined to a four-dimensional
brane localized at the
orbifold fixed point at \il{x^5=0}.

Within this geometry we consider 
a five-dimensional scalar field $\Phi(x^\mu, x^5)$  and a
corresponding five-dimensional action ${\cal S}$ of the form
\beq\label{eq:5Daction}
 {\cal S} ~=~ \int d^4x\,dx^5 \left[\mathcal{L}_{\rm bulk}(\Phi)
+ \delta(x^5)\mathcal{L}_{\rm brane}(\psi_i,\Phi)\right] 
\eeq
where $\psi_i$ are fields confined to the SM brane. 
In general, our bulk action will contain
generic kinetic and mass terms of the form 
\beq\label{eq:generalLbulk}
\mathcal{L}_{\rm bulk}(\Phi) ~=~ \frac{1}{2}\p_M\Phi^*\p^M\Phi - \frac{1}{2}M^2\abs{\Phi}^2~.
\eeq
However, many well-motivated 
scalars such as moduli and axions have a primordial shift 
symmetry under which \il{\Phi\to \Phi+c} for arbitrary constants $c$. 
With this motivation in mind, we shall henceforth assume that \il{M=0}.

The brane Lagrangian $\mathcal{L}_{\rm brane}$, on the other hand, 
includes the SM Lagrangian in addition 
to interactions between $\Phi$ and the SM fields. In general, these interactions can
open up decay channels from $\Phi$ to SM states.   However, such interactions can also lead
an {\it effective}\/ four-dimensional mass for $\Phi$ on the brane.
Such masses are typically generated as the result of some brane-localized dynamics 
(such as a brane-localized phase transition) 
which occurs at some time during the cosmological evolution on the brane
and which explicitly breaks the shift symmetry. 
For this reason, we shall allow our brane Lagrangian to include
a time-dependent effective brane-mass term of the form
\beq\label{eq:Lbrane}
{1\over {\cal V}}\, \mathcal{L}_{\rm brane}(\Phi) ~=~ -  \frac{1}{2}m^2(t) \abs{\Phi}^2 + ...  \ 
\eeq
where \il{{\cal V}\equiv 2\pi R} is the compactification volume.
The behavior of $m(t)$ is ultimately determined by the details of the non-perturbative dynamics on 
the brane and thus highly model-dependent. 

This information alone is sufficient to allow us to determine
the effective four-dimensional theory that emerges
upon compactification.
Assuming that $\Phi$ has even parity under the orbifold action \il{x^5\to -x^5},
we may perform a Kaluza-Klein expansion of the form
\beq
\Phi(x^{\mu},x^5) ~=~ \frac{1}{\sqrt{\cal V}}\sum_{k=0}^{\infty}r_k
\phi_k(x^{\mu})\cos\left(\frac{kx^5}{R}\right) \ 
\label{eq:orbifoldmodeexpansion}
\eeq
with normalization constants \il{r_0=1} and \il{r_{k} =\sqrt{2}} for all \il{k>0}, as appropriate
for a $S^1/\IZ_2$ compactification.
Inserting this into our action and integrating over $x^5$ we obtain
an effective four-dimensional action of the form 
\begin{align}
\mathcal{L}_{\rm eff} ~=~ \sum_{k=0}^{\infty}\left[\frac{1}{2}(\p_{\mu}\phi_k)^2
- {1\over 2}\sum_{\ell=0}^{\infty}\phi_k\mathcal{M}^2_{k\ell}(t)\phi_\ell\right] ~+~...~~~ 
\label{eq:4Dlagrangian}
\end{align}
where our time-dependent mass matrix $\calM_{k\ell}^2$ is given by 
\beqn
     \calM_{k\ell}^2 &=& k\ell \, \delta_{k\ell} \, M_c^2 ~+~
       r_k r_\ell\, m^2 (t)~\nonumber\\
&=& m^2(t)
\begin{bmatrix}
1  & \sqrt{2} & \sqrt{2} & \cdots \\
\sqrt{2} & ~~2 + \frac{M_c^2}{m^2(t)} & 2 & \cdots \\
\sqrt{2} & 2 & ~~2 + \frac{4M_c^2}{m^2(t)}~ & \cdots \\
\vdots & \vdots & \vdots & \ddots 
\end{bmatrix} ~~~~~
\label{kmassmatrix}
\eeqn
with \il{M_c\equiv 1/R}.
The first term in the top line of Eq.~(\ref{kmassmatrix}) represents the expected KK contribution
from the physics in the bulk,
while the second term reflects the contribution from physics on the brane
which breaks the translational symmetry in the $x^5$-direction and thereby
introduces a mixing between the different KK modes.

Given the Lagrangian in Eq.~(\ref{eq:4Dlagrangian}), 
it follows that in a flat Friedman-Robertson-Walker (FRW) 
spacetime the zero-momentum modes of the KK scalars $\phi_k$ obey the coupled equations of motion
\beq
\ddot{\phi}_k + 3H(t)\dot{\phi}_k + \sum_{\ell = 0}^{\infty}\mathcal{M}^2_{k\ell}(t)\phi_\ell ~=~ 0 ~
\label{eq:KKequationsofmotion}
\eeq
where \il{H(t)\equiv \kappa/(3t)} is the Hubble parameter.
Note that \il{\kappa=2} and \il{\kappa=3/2} for matter- and radiation-dominated epochs, respectively.
If $\calM^2_{k\ell}$ were constant as a function of time, it would be possible to 
pass to a mass-eigenstate basis \il{\lbrace\phi_\lambda\rbrace} 
in which the different differential equations in Eq.~(\ref{eq:KKequationsofmotion})
would decouple.  Unfortunately, the time-dependence of $m(t)$ leads to a time-dependence
for $\calM^2_{k\ell}$.    This time-dependence induces an unavoidable mixing between the different
modes, since even the mass-eigenstate basis is continually changing.
It is this feature which underpins the non-trivial dynamics we shall be studying in this paper.
Likewise, at any moment in time, the total energy density associated 
with our system of KK modes is given by
\beq
\rho ~=~ {1\over 2}\sum_k \dot\phi_k^2 + 
         {1\over 2}\sum_{k,\ell} \phi_k \calM_{k\ell}^2 \phi_k~.
\label{energydensity}
\eeq
If the mass matrix $\calM_{k\ell}^2$ were time-independent, it would be
possible to rewrite this energy density in the mass-eigenstate basis 
as the sum of individual contributions: 
\beq
            \rho  ~=~ \sum_\lambda \rho_\lambda
\label{energydecomp}
\eeq
where
\beq 
	\rho_\lambda ~=~ {1\over 2}\left(
	\dot\phi_\lambda^2 + \lambda^2 \phi_\lambda^2 \right)~.
\label{rholambda}
\eeq
Unfortunately, once again, the time-dependence of the mass matrix renders such a decomposition impossible.
Thus, we may consider the total energy density in Eq.~(\ref{energydensity}) as receiving contributions
from individual mass eigenstates as in Eq.~(\ref{energydecomp}) only during those periods of cosmological evolution
for which the mass matrix might not be changing with time. 

In general, we shall assume that the mass $m(t)$ is generated through some sort of phase
transition on the brane whose occurrence is centered at some time $t_G$ in cosmological history  
and which requires a duration $\Delta_G$ over which to unfold.
We shall therefore assume that \il{m(t)=0}
long before the phase transition (\ie, for \il{t\ll t_G}), 
while we shall assume that $m(t)$ is given by some fixed mass $\overline{m}$ long after the phase
transition (\ie, for \il{t\gg t_G}).
The precise details of how $m(t)$ evolves from $0$ to $\overline{m}$ clearly depend on the particular
structure of the phase transition in question, but we shall find that the most important feature of
this function will be its ``width'' $\Delta_G$. 
For concreteness, we shall therefore adopt the function~\cite{TwoTimescales} 
\beq
m(t) ~=~
\frac{1}{2}\,\mbar\, \left\{1 + 
\text{erf}\left[\frac{1}{\sqrt{2}\delta_G}\log\left(\frac{t}{t_G}\right)\right]\right\} \ ,
\label{sigmoid}
\eeq
where $\mbar$ is the mass at asymptotically late times, where $t_G$ is the central time
defined by \il{m(t_G)=\mbar/2}, and where `erf' denotes the so-called ``error'' function. 
In Eq.~(\ref{sigmoid}), the quantity \il{\delta_G\in\left[0,1\right]} is a 
dimensionless parameter describing the ``width'' of the phase transition,
with  \il{\delta_G=0} corresponding to an essentially
instantaneous phase transition and \il{\delta_G=1} describing a phase transition
which proceeds as slowly as possible.
Given the dimensionless parameter $\delta_G$, the physical timescale $\Delta_G$ over which
the phase transition occurs is then given by~\cite{TwoTimescales}
\beq
\Delta_G ~\equiv~ \sqrt{2\pi}\delta_G t_G \ .
\label{bigwidth}
\eeq
Indeed, given Eq.~(\ref{sigmoid}), we may view the transition from \il{m=0} to \il{m=\mbar} as 
occurring over a ``width'' $\Delta_G$ logarithmically centered at $t_G$. 

This choice of function is discussed more fully in Ref.~\cite{TwoTimescales}.  Indeed, it is 
shown in Ref.~\cite{TwoTimescales} that this function models the time-dependence of a variety 
of phase transitions extremely well in the neighborhood of $t_G$, where the mass function is 
changing most rapidly.   This includes the instanton-induced phase transition 
that gives mass to the QCD axion.   Of course, as discussed in 
Refs.~\cite{WantzShellard1,WantzShellard2,diCortonaQCDAxion,BorsanyiQCDAxion},
the actual QCD phase transition leads to a mass function $m(t)$ which exhibits a power-law 
dependence on $t$ at early times $t\ll t_G$, and which is then often taken to be followed by 
one or more slope discontinuities near \il{t\sim t_G} before becoming a constant for \il{t\gg t_G}.    
By contrast, our functional form for $m(t)$ is designed to interpolate smoothly between fixed 
early- and late-time values in such a way that there is an adjustable and easily identifiable 
``width'' $\Delta_G$.  Despite these differences, however, it is shown in Ref.~\cite{TwoTimescales} 
that our functional form models the QCD functional form extremely well near \il{t\sim t_G}.
 
Of course, many other choices are possible for the particular functional form of 
$m(t)$, with different choices being more or less suitable in different contexts.
However, none of the qualitative results of this paper will ultimately depend on 
these specific details.

In general, we may choose to consider phase transitions with any \il{\delta_G\leq 1}.
Indeed, as discussed in Ref.~\cite{TwoTimescales}, situations with \il{\delta_G > 1} suffer
from an unphysical boundary artifact near \il{t\approx 0}, namely the emergence of a ``fake'' 
mass-generating phase transition whose width tends to decrease as $\delta_G$ increases beyond \il{\delta_G=1}.
However, when \il{\mbar/M_c\gg 1}, it turns out that an additional
boundary effect emerges which restricts $\delta_G$ even more strongly, ultimately imposing an upper limit on $\delta_G$ which
is less than $1$ and which depends on $\mbar/M_c$.   
(Specifically, as $\delta_G$ increases for \il{\mbar/M_c\gg 1}, the effective ``mixing angle'' $\theta$ between our ground
state and the first excited state starts changing, evolving from \il{\theta=0} to \il{\theta=\overline{\theta}} 
more and more rapidly and at earlier and earlier
times.  Ultimately we reach a new critical value of $\delta_G$ beyond which the $\theta$-width $\Delta_\theta$ in
Eq.~(4.3) of Ref.~\cite{TwoTimescales} actually starts to {\it decrease}\/ when $\delta_G$ is
increased, giving the appearance of a new ``fake'' 
phase transition at early times near $t\approx 0$.)    
Avoiding this unphysical behavior then 
imposes an additional upper bound on $\delta_G$
which we can evaluate numerically as a function of $\mbar/M_c$ and $N$.
In this paper we shall therefore restrict our attention to situations with \il{\delta_G\leq 1} but also impose
this additional constraint as appropriate for large $\mbar/M_c$. 

In this context, we note that
theories with extra spacetime dimensions must generally obey additional ``normalcy-temperature'' 
constraints~\cite{ADD,ADDPheno} 
which are more easily satisfied within cosmological low-temperature reheating (LTR) scenarios 
in which inflation occurs late and reheating occurs at temperatures as low as 
\il{T_{\rm RH}\sim \mathcal{O}(\text{MeV})}~\cite{KawasakiLTR}. 
If we assume the dynamics that generate $m(t)$ are at higher scales than $T_{\rm RH}$, then our 
mass-generating phase transition can be assumed to occur during 
an inflaton-dominated --- \ie, a matter-dominated --- epoch.
However this assumption will not play a significant role in our analysis, and the primary
results of this paper will hold regardless of the specific cosmological timeline assumed.

Given the differential equations of motion in Eq.~(\ref{eq:KKequationsofmotion}),
the only ingredients of our model remaining to be specified 
are the initial conditions on each field mode $\phi_k$.   
While in principle many possibilities 
exist, 
the existence of the shift symmetry \il{\Phi \rightarrow \Phi + c} prior to mass
generation on the brane suggests 
that $\Phi$ might have an arbitrary fixed displacement at early times \il{t_i\ll t_G},
with a fixed non-zero vacuum expectation value (VEV) \il{\langle \Phi\rangle\neq 0}.
Upon KK reduction, this corresponds to initial conditions given by
\begin{align}\label{eq:initialconditions}
\phi_k(t_i) ~&=~ \expt{\phi_0} \delta_{0k} \nn \\
\dot{\phi}_k(t_i) ~&=~ 0 \ ,
\end{align}
where \il{\expt{\phi_0}\equiv \sqrt{2\pi R}\expt{\Phi}}.
We shall therefore take these to be the initial 
conditions for our differential equations (\ref{eq:KKequationsofmotion}) 
in what follows.
It is important to note that these initial conditions 
have several additional advantages beyond their natural origins outlined above.
First, as long as the initial time $t_i$ is sufficiently prior to $t_G$, our ensemble of KK modes
will have zero energy.   Thus, as we shall see, {\it all energy accrued by our KK system is 
solely the result of the phase transition on the brane}\/. 
Moreover, these initial conditions, being essentially static,
free our system and its subsequent dynamics
from all details concerning
the generation of the initial VEV \il{\langle \Phi\rangle}.
As a result, the time at which the initial VEV \il{\langle \Phi\rangle} 
is generated is arbitrary and we need not concern ourselves with its origins. 
Indeed, as we shall see, the precise value of $t_i$ will not affect our analysis, or any of our conclusions.
Finally, we remark that while $\expt{\phi_0}$ sets an overall energy scale, this scale
will be irrelevant for our purposes since the equations of motion are linear and
our main interest will be on relative {\it comparisons}\/ between energy scales 
rather than their absolute magnitudes.

Finally, let us discuss the masses of the individual KK modes in this model.
Even though our overall brane mass $m(t)$ is continually changing in time,
diagonalizing the matrix in Eq.~(\ref{kmassmatrix}) at any moment in time
leads directly to the corresponding instantaneous (mass)$^2$ eigenvalues $\lambda_k^2$.
While no closed-form analytical expressions for these eigenvalues
exist, they may be readily approximated in the \il{m/M_c\ll 1} and \il{m/M_c\gg 1} limits.
In the \il{m/M_c\ll 1} limit,
we find
\beq
      \lambda_k ~\approx~ \begin{cases}
              m  &   k=0\cr
             k M_c & k>0~.
           \end{cases}
\label{eq:unmixedmassspectrum}
\eeq
There are two ways to interpret this result.
If we imagine holding $M_c$ fixed, this limit corresponds to taking $m$
extremely small.
This renders the brane irrelevant in the KK mass decomposition, and
indeed 
the results in Eq.~(\ref{eq:unmixedmassspectrum})
are then the eigenvalues expected from a straightforward compactification on a circle.
Alternatively, if we imagine holding $m$ fixed, we see that this limit corresponds
to taking $M_c$ extremely large.   
We then have only a single light mode with mass \il{\lambda_0\approx m}, which is nothing
but the four-dimensional limit.
By contrast, for \il{m/M_c\gg 1},
 the mixing between the modes is maximized and our corresponding eigenvalues are given 
by~\cite{DDGAxions,DDM1} 
\beq
      \lambda_k ~\approx~ (k+1/2) \,M_c~.
\label{kmixedlimit}
\eeq
It is remarkable that the effect of the brane mass in this limit is entirely $m$-independent, and merely amounts
to shifting our eigenvalues by $M_c/2$.   Indeed, these shifted eigenvalues are those that would have emerged from 
an {\it anti}\/-periodic compactification on a circle.
Finally, for intermediate values of $m/M_c$, we find that our eigenvalues 
$\lambda_k$
tend to follow Eq.~(\ref{kmixedlimit}) for \il{k\ll \pi m^2/M_c^2} and   
Eq.~(\ref{eq:unmixedmassspectrum}) for \il{k\gg \pi m^2/M_c^2},
with $\lambda_k$ taking values that smoothly evolve between these two extremes for other values
of $k$.

It is important to stress that this behavior for the eigenvalues emerges
only in the limit in which we consider the full KK tower with its infinite complement
of KK modes, \il{k=0,1,...,\infty}.
By contrast, if we truncate the mass matrix in Eq.~(\ref{kmassmatrix}) to its first
$N$ rows and columns so that \il{0 \leq \lbrace k,\ell\rbrace \leq N-1}, 
our general expectations described above continue to apply
only for the eigenvalues \il{\lambda_0, ..., \lambda_{N-2}}.
However, for the highest eigenvalue $\lambda_{N-1}$, we find that the above
expectations continue to apply only for \il{m^2/M_c^2 \lsim {\cal O}(\sqrt{N})}.
By contrast, for \il{N\gg 1} and 
for \il{m^2/M_c^2 \gsim {\cal O}(\sqrt{N})},
we instead find that
\beq
     \lambda_{N-1} ~\approx~  \sqrt{2 N} \, m ~.
\label{artifact}
\eeq
The normalized eigenvalue $\lambda_{N-1}/M_c$ thus diverges as \il{m/M_c\to \infty}.

\begin{figure}
\includegraphics[keepaspectratio, width=0.5\textwidth]{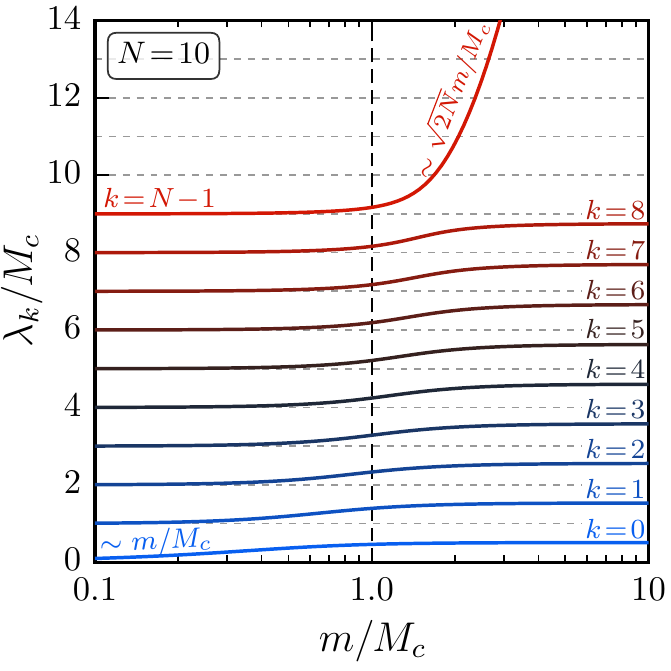}
\caption{The mass spectrum $\lambda_k$ for \il{N=10}, plotted as a function of $m/M_c$.
When \il{\mbar\ll M_c}, 
our mass eigenvalues follow the simple form given in Eq.~(\ref{eq:unmixedmassspectrum}).
By contrast, when
\il{\mbar\gg M_c},
all but the highest mass eigenvalue shift upwards
by $M_c/2$, while the highest mass eigenvalue diverges according
to Eq.~(\ref{artifact}) for \il{m^2/M_c^2\gsim {\cal O}(\sqrt{N})}.
As discussed in the text, this anomalous behavior
of the highest mass eigenvalue is an artifact 
of having truncated our KK tower to only a finite number of states, and 
disappears as \il{N\to \infty}.}
\label{fig:massspectrum}
\end{figure}
 
The behavior of the eigenvalues $\lambda_k$ is illustrated in Fig.~\ref{fig:massspectrum},
where we have plotted the values of $\lambda_{k=0,...,9}$
as functions of $m/M_c$ for \il{N=10}.
As is clear from Fig.~\ref{fig:massspectrum}, the somewhat anomalous behavior of 
$\lambda_{N-1}$ is ultimately an artifact of truncating 
our KK tower to its first $N$ values.
However, as $N$ increases, this anomalous behavior of $\lambda_{N-1}$ 
sets in only at increasingly 
large values of $m/M_c$.
Thus, in the \il{N\to\infty} limit, this anomalous behavior disappears
altogether, as expected.

In our model the quantity $m$ is  
generally time-dependent, rising from zero to some final, fixed late-time value $\mbar$
as the result of our mass-generating phase transition on the brane.
The corresponding eigenvalues $\lambda_k^2$ are thus set  
by Eq.~(\ref{eq:unmixedmassspectrum}) with \il{m=0} at early times, but 
subsequently evolve towards late-time results which are governed 
by the magnitude of the late-time ratio $\mbar/M_c$.
Meanwhile, each of the corresponding KK fields $\phi_k$ is evolving 
according to Eq.~(\ref{eq:KKequationsofmotion}), with the non-diagonal
mass matrix $\calM^2_{k\ell}$ inducing a time-dependent mixing between the KK modes
and even between their instantaneous mass eigenstates.
This, coupled with the continuously decreasing Hubble friction term $H(t)$ in 
Eq.~(\ref{eq:KKequationsofmotion}), leads to a highly non-trivial dynamical system,
with energy slowly being introduced into the system as the result of the mass-generating
phase transition while it is simultaneously and continually redistributed across the different
modes and dissipated from each mode.
Understanding this energy flow is critical if we are to properly
understand the cosmological implications of such KK scalar towers in the presence
of mass-generating phase transitions in the early universe. 
This is therefore the main task to which we now turn.

\FloatBarrier
\section{\texorpdfstring{$N=1$}{N=1}:  ~Mapping out the 4D Limit \label{sec:FourDimensionalLimit}}

We begin our analysis by focusing on the simplest case:  that of only one scalar field, \ie, the case
with \il{N=1}.   This case lacks any dependence on $M_c$, and therefore can be 
considered equivalent to  a four-dimensional limit in which \il{M_c\gg m(t)} 
at all times (so that all KK modes with \il{k\geq 1} effectively decouple from the problem).
This case of a single scalar field $\phi(t)$ undergoing a mass-generating phase transition
has been studied extensively in the literature (see, \eg, Refs.~\cite{TurnerEffect,UpdatedAxionAbundanceCalc}), 
but only in certain ``adiabatic'' or ``abrupt'' limits.
In this section, by contrast, we shall provide a complete mapping of the entire relevant 
parameter space.

In general, given the equation of motion in Eq.~(\ref{eq:KKequationsofmotion}),
we see that the field $\phi(t)$ simply follows the trajectory of a damped harmonic oscillator with time-dependent
critical damping coefficient \il{\zeta \equiv 3H/2m}.  Such an oscillator is overdamped if \il{\zeta>1} 
and only experiences oscillations 
in the underdamped regime with \il{\zeta<1}.
Since \il{H(t)\sim \kappa/(3t)} 
is monotonically decreasing while $m(t)$ is monotonically increasing,  
we see that $\zeta$ is constantly decreasing.
As a result,
the field $\phi(t)$ is necessarily overdamped at times \il{t \ll t_G} 
and does not begin to undergo coherent oscillations until \il{\zeta = 1}.
We shall let $t_\zeta$ denote this time at which such coherent oscillations begin.
In general, we know that $t_\zeta$ cannot come too much earlier than $t_G$ 
(\ie, we cannot have \il{t_G-t_\zeta \gg \Delta_G}) since our
field $\phi$ is presumed massless prior to the onset of the phase transition.
Likewise, if $t_{\zeta}$ occurs long after the phase transition has 
already occurred (\ie, if \il{t_\zeta - t_G\gg \Delta_G}), the specific details of
the phase transition such as its shape or width will have a negligible effect 
on the resulting dynamics and on the corresponding energy 
density $\rho$. However, if \il{|t_{\zeta}-t_G|\lsim \Delta_G},
the specific properties of the phase transition can be significant.

Two approximations are typically used in the literature in order to estimate the effects of
the phase transition  within this third regime.
Both rest upon considering extreme limits of the phase-transition width $\Delta_G$. 
If the width is large enough so that \il{\dot m/m \ll m} at all times during the  
field oscillations (\ie, at all times after $t_\zeta$),
then the field undergoes many oscillations during the phase transition. 
This is the so-called \emph{adiabatic approximation}:  the fields remain virialized during the phase transition and we 
can approximate the energy density by
\beq
\rho_{\rm 4D}(t) ~\approx~ 
\frac{1}{2}\, \langle\phi_0\rangle^2 \, m(t_{\zeta}) \, m(t)\, \left[\frac{a(t_{\zeta})}{a(t)}\right]^3 
\label{eq:adiabaticapproximation}
\eeq
where \il{a(t)\sim t^{\kappa/3}} is the cosmological scale factor.
By contrast, if the phase transition width $\Delta_G$ is extremely small, we are in the so-called
\emph{abrupt approximation}:   our assumption that
\il{|t_{\zeta}-t_G|\lsim \Delta_G}
forces $t_\zeta$ and $t_G$ to coincide, and the phase transition happens so rapidly 
that the field $\phi(t)$ retains its initial value \il{\langle\phi_0\rangle} 
until $t_G$ (or equivalently $t_\zeta$), at which point it immediately begins oscillating coherently.
The evolution of the energy density is then easily obtained by solving 
Eq.~\eqref{eq:KKequationsofmotion} with \il{N=1} analytically.   This yields the exact solution
\beq
\rho_{\rm 4D}(t) ~=~ \frac{\pi^2}{8}\, \langle\phi_0\rangle^2\, \mbar^4\, \frac{t_G^{\kappa+1}}{t^{\kappa-1}}\, 
\left[B_1^2(t) +B_2^2(t)\right] \ ,
\label{eq:abruptapproximation}
\eeq
where 
\beqn
B_1(t) &\equiv& J_{\kappa_+}(\mbar t_G)Y_{\kappa_-}(\mbar t) - 
J_{\kappa_-}(\mbar t)Y_{\kappa_+}(\mbar t_G) \nonumber \\
B_2(t) &\equiv& J_{\kappa_+}(\mbar t)Y_{\kappa_+}(\mbar t_G) - 
J_{\kappa_+}(\mbar t_G)Y_{\kappa_+}(\mbar t) \nonumber\\
\eeqn
and where we have defined \il{\kappa_{\pm} \equiv (\kappa \pm 1)/2}. 
Note that in both Eq.~(\ref{eq:adiabaticapproximation}) 
and Eq.~(\ref{eq:abruptapproximation}),
the quantity $t$ is measured against  the same cosmological clock
that measures $t_G$. 

While the approximations above are applicable in two limits of parameter space, it is 
important to understand the behavior of the late-time energy density over the entire parameter space.
In particular, although the mass-generating phase transition generally pumps energy into the system,
and although this energy density is ultimately dissipated by the Hubble friction that 
slowly damps the resulting field oscillations,
it turns out that there can be an additional source of dissipation
  and therefore an additional source of suppression of the
late-time energy density $\rhobar_\fd$ compared to our usual expectations based on the abrupt approximation.
This arises if the corresponding field is undergoing oscillations 
 {\it during}\/ the phase transition, while the mass of the field is changing appreciably.

\begin{figure*}[t]
\includegraphics[keepaspectratio, width=0.48\textwidth]{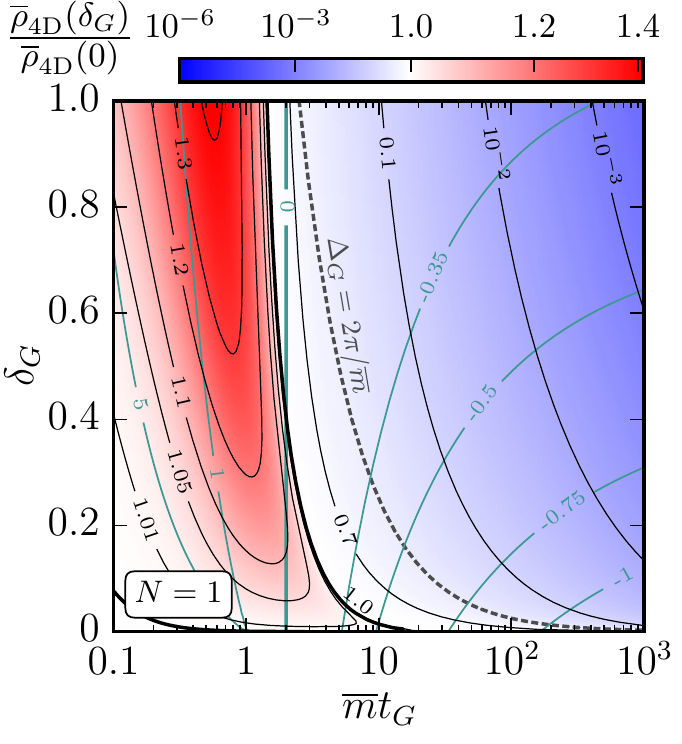}
\hskip 0.2 truein
\includegraphics[keepaspectratio, width=0.48\textwidth]{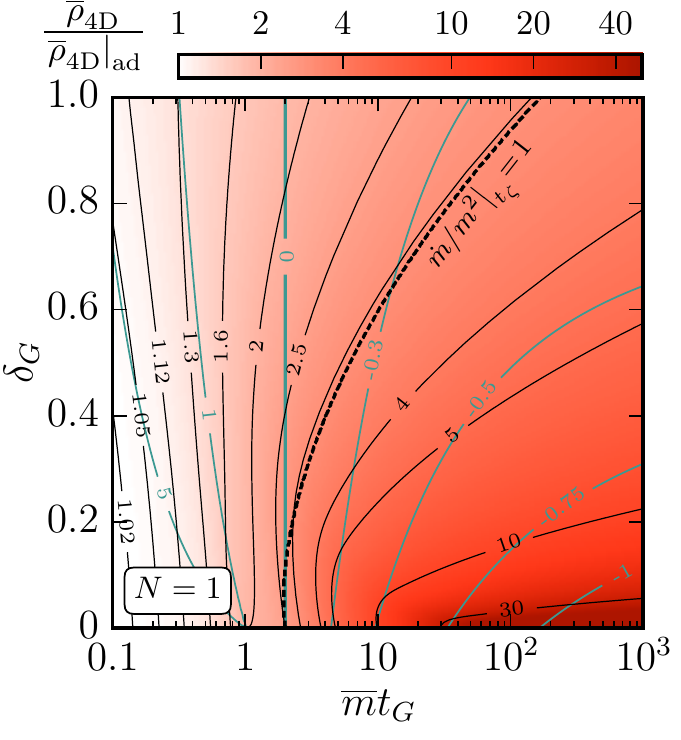}
\caption{The late-time energy density $\rhobar_\protect\fd$, plotted within the \il{(\mbar t_G, \delta_G)} plane
and normalized to the value it would have had in the abrupt approximation (left panel)
or adiabatic approximation (right panel).
In each case the colors and black lines indicate the contours of the normalized $\rhobar_\fd$ 
(\ie, the degrees to which the true late-time energy density is enhanced or suppressed
relative to its approximated value),
while the blue lines indicate the corresponding contours of \il{(t_\zeta-t_G)/\Delta_G}.
As additional relevant guidelines,  
the dashed line in the left panel 
indicates the contour
along which \il{\Delta_G=2\pi/\mbar},
while in the right panel 
it indicates the contour along which \il{\dot{m}(t_{\zeta}) = m^2(t_{\zeta})}.
In general, the energy densities in regions above or to the left of these
guidelines 
in the left (right) panel tend to obey the abrupt (adiabatic) approximation 
more strongly than those below or to the right.  
We also find that neither approximation holds 
across the majority of the parameter space shown, with the true
values of the late-time energy density $\rhobar_\fd$ often experiencing significant
suppressions (blue regions) or enhancements (red regions) as compared with the 
approximated expectations. }
\label{fig:singlefielddensityplot}
\end{figure*}

It is easy to identify those regions of parameter space for which this will be the case.
In general, there are two criteria that must be satisfied 
if the scalar field is to undergo at least one oscillation 
while the mass of the field is changing appreciably.
First, the scalar field must indeed be undergoing oscillations at some point during the phase
transition --- \ie, our system must be in the underdamped regime before the mass of our field reaches $\mbar$.
We thus must have \il{3H \lsim 2 \mbar}, or equivalently \il{\mbar t_G\gsim  \kappa/2},
where we have taken $t_G$ as a rough benchmark time for the phase transition.  
Second, 
in order to ensure that at least one or more field oscillations
occur during the interval over which the mass is changing,
we require that the timescale associated with the field oscillations 
be shorter than the phase-transition timescale associated with
the changing mass.
The former timescale is generally given by $2\pi/m$ and thus decreases
throughout the phase transition, ultimately reaching a minimum value $2\pi/\mbar$,
while the latter timescale is nothing but $\Delta_G$.   
As a rough benchmark, our phase transition will therefore include at least one oscillation
so long as \il{\Delta_G \gsim 2\pi/\mbar}, or equivalently
\il{\mbar t_G \gsim \sqrt{2\pi}/\delta_G}.
Combining our two conditions, we thus find that an additional suppression
of the late-time energy density compared with what we might expect based on the abrupt approximation
will occur as long as our system satisfies 
the rough benchmark criterion
\beq
            \mbar t_G ~\gsim~ 
         {\rm max} \left\lbrace  {\sqrt{2\pi}\over \delta_G}\, ,\, {\kappa\over 2} \right\rbrace~.
\eeq
Note that for a matter- or radiation-dominated background cosmology, the underdamped condition will automatically
be satisfied whenever the oscillation condition is satisfied. 

In the left panel of Fig.~\ref{fig:singlefielddensityplot}, we 
plot the exact numerical value of the late-time
energy-density ratio $\rhobar_\fd(\delta_G)/\rhobar_\fd(0)$ within 
the ($\mbar t_G$,$\delta_G$) plane.
For reference, the benchmark contour with
\il{\Delta_G=2\pi/\mbar} is also plotted.
In general, 
the regions above and to the right
of this contour (\ie, regions with \il{\mbar t_G \gg 1})
indeed experience suppressions which grow increasingly
severe compared with what we might expect from the abrupt approximation.
As indicated in Fig.~\ref{fig:singlefielddensityplot}, these are regions
with \il{t_\zeta > t_G}.
Surprisingly, however, we note that there is also a region in which
the late-time energy density is {\it enhanced}\/ compared with its abrupt expectation.
This region arises only for \il{\mbar t_G\sim {\cal O}(1)}, and typically has \il{t_\zeta-t_G \lsim
\Delta_G}.
Like the corresponding suppression,
this too is a feature of considering
a non-zero timescale for the phase transition, occurring only when \il{\delta_G >0}.
Finally, for \il{\mbar t_G \ll 1} or for \il{\delta_G \ll 1},
we see that our late-time energy density is neither enhanced 
nor suppressed.  Indeed, these are the regions in which the abrupt approximation
applies.
 
In the right panel of Fig.~\ref{fig:singlefielddensityplot}, 
we plot the same late-time energy density $\rhobar_\fd(\delta_G)$, only now normalized  
to what we might expect from the {\it adiabatic}\/ approximation.
In this case, the contour 
with \il{\dot m/m^2=1} at \il{t=t_\zeta} is shown
as a relevant guideline.
Unlike the comparison to the abrupt
approximation, in this case
we see that the late-time energy density $\rhobar_\fd(\delta_G)$
is {\it enhanced}\/ throughout the entire region
with \il{\delta_G >0}.   As expected, this enhancement is relatively mild  
in regions to the left of the
\il{\dot m/m^2=1} contour but grows increasingly severe below and to the right.
Once again, it is only for \il{\mbar t_G \ll 1} that the adiabatic approximation
appears to hold.

We see, then, that our usual expectations based on the abrupt or adiabatic approximations
apply only in relatively small regions of the full \il{(\mbar t_G,\delta_G)} parameter space, the
former consisting of the 
regions with \il{\mbar t_G \ll 1} or \il{\delta_G \ll 1} and the latter consisting of
the region with \il{\mbar t_G \ll 1}.
Interestingly, we see that {\it either}\/ approximation yields the same 
(reliable) result for \il{\mbar t_G\ll 1}, regardless of the value of $\delta_G$.
In all other regions, however, we see that these standard approximations break down ---  often
significantly ---  with the adiabatic approximation tending to underestimate the true value
and the abrupt approximation either under- or over-estimating the true value, depending
on the particular values of $\mbar t_G$ and $\delta_G$.

One of the advantages of the abrupt and/or adiabatic approximations is that they 
provide relatively simple, {\it analytical}\/ expressions for the late-time energy
density $\rhobar_\fd(\delta_G)$.
Unfortunately, as we have seen, these expressions are accurate
only within relatively narrow slices of the full parameter space.
Given this deficiency, we now offer  
two approximate analytical expressions for $\rhobar_\fd(\delta_G)/\rhobar_\fd(0)$ which
together describe its numerical values fairly accurately across the entire parameter space shown in
Fig.~\ref{fig:singlefielddensityplot}. 
Subsequent use of the analytical expression in 
Eq.~(\ref{eq:adiabaticapproximation})
then allows us to isolate $\rhobar_\fd(\delta_G)$.

We begin by noting that within the 
suppressed region 
in the left panel of 
Fig.~\ref{fig:singlefielddensityplot}, 
the contours of $\rhobar_\fd(\delta_G)/\rhobar_\fd(0)$ 
roughly follow the same slope as the 
contour with 
\il{\Delta_G=2\pi/\mbar}.  Indeed, for
\il{\delta_G\lsim 0.3}, 
one finds the approximate power-law scaling behavior
\beq
      {\rhobar_\fd(\delta_G) \over \rhobar_\fd(0)}  ~\sim~ { 1\over \mbar t_G \delta_G}~.
\label{fdscaling}
\eeq
Of course, as we move closer to the \il{\delta_G=1} boundary, 
the contours deviate from this scaling behavior and the suppression grows increasingly severe. 
Nevertheless, we find an approximate expression 
\beq\label{eq:4Dsuppressedformula}
\frac{\rhobar_{\rm 4D}(\delta_G)}{\rhobar_{\rm 4D}(0)} ~\approx~ 
\frac{(\mbar t_G \delta_G)^{-0.92}}{1 + \left[(\mbar t_G)^{1/4}-3/2\right]\delta_G^2}~ 
\eeq
which holds to within \il{\pm 20\%} across
the entire suppressed region and which approximately reduces to the power-law result 
in Eq.~(\ref{fdscaling}) for very small $\delta_G$.

By contrast, turning to the enhanced region near \il{\mbar \sim 1/t_G},
we see the maximum of this enhancement always occurs at \il{\delta_G=1}. 
Indeed, we find that we can approximate this enhancement analytically in
the Gaussian form
\beq\label{eq:4Denhancedformula}
\frac{\rhobar_{\rm 4D}(\delta_G)}{\rhobar_{\rm 4D}(0)} ~\approx~ 
1 + 0.42 \exp\left[-\ve{v}^T\!\!\begin{pmatrix}1.64 & 0.68 \\ 0.68 & 0.67\end{pmatrix}\!\ve{v}\right] \ ,
\eeq
where \il{\ve{v}\equiv \left[\, \log(\mbar t_G/0.54) , \, \log(\delta_G)\,\right]^T}. 
In this case, the error associated with this approximation
is bounded to lie between $-10\%$ and $5\%$ throughout
the non-suppressed portion of the parameter space shown.
Thus, with the analytical expressions in Eq.~\eqref{eq:4Dsuppressedformula} and 
Eq.~\eqref{eq:4Denhancedformula},
we now cover the entire single-scalar parameter space.
We shall nevertheless continue to use our exact results throughout the rest of this paper.

In Fig.~\ref{fig:singlefielddensityplot} 
we have presented our results for the late-time energy density $\rhobar_\fd$ 
as fractions of the values this quantity would have had in either the abrupt or adiabatic
approximations.
These results therefore enabled us to understand the effects that can accrue beyond
the regions in which these approximations are valid.
However, it is also important to consider the {\it absolute}\/ magnitudes of $\rhobar_\fd$
that are generated throughout our \il{(\mbar t_G, \delta_G)} parameter space.
These can be trivially obtained at each point in parameter space by multiplying the results in 
the left panel of Fig.~\ref{fig:singlefielddensityplot} 
by the corresponding values in 
Eq.~(\ref{eq:abruptapproximation})
[or equivalently the results in the right panel of 
Fig.~\ref{fig:singlefielddensityplot} 
by the corresponding values in 
Eq.~(\ref{eq:adiabaticapproximation})].

\begin{figure}[b]
\includegraphics[keepaspectratio, width=0.5\textwidth]{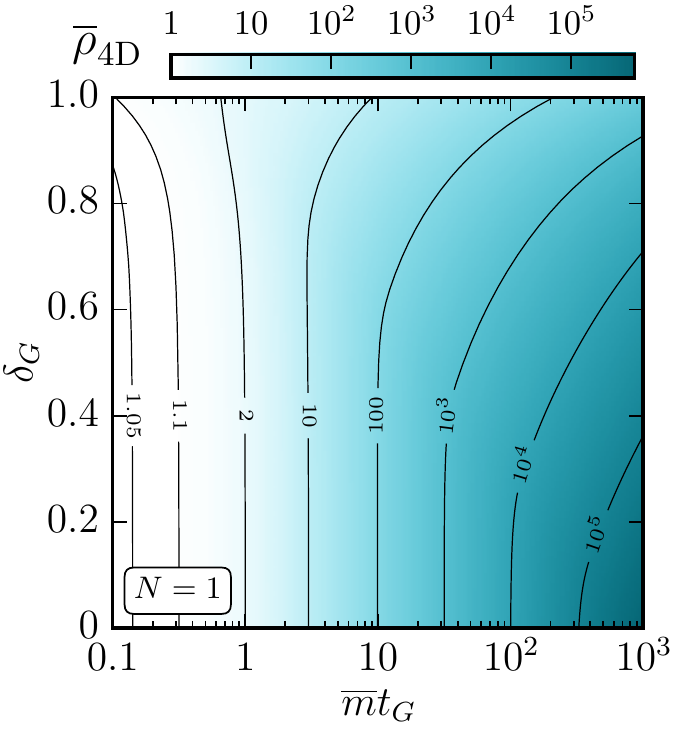}
\caption{The {\it absolute}\/ late-time energy density $\rhobar_\fd$, plotted in units
of \il{\half \langle \phi\rangle^2 t^{-2}} within the \il{(\mbar t_G,\delta_G)} plane.
Note that our late-time energy density $\rhobar_\fd$
is generally independent of $\delta_G$ for \il{\lbrace \mbar t_G\lsim 10^3, \delta_G\lsim 0.5\rbrace}
and scales roughly as $(\mbar t_G)^2$ within this region.}
\label{fig:fdabs}
\end{figure}

The results are shown in Fig.~\ref{fig:fdabs}, where we plot the absolute
magnitude of $\rhobar_\fd$ in units of \il{\half \langle \phi\rangle^2 t^{-2}}.
Once again, $t$ refers to universal time, measured against the same cosmological clock as $t_G$.
Thus, for example, if our late-time mass is chosen to be \il{\mbar t_G=10} and our mass-generating phase transition width
is chosen to be \il{\delta_G=0.4}, we find  
from Fig.~\ref{fig:fdabs} that our resulting late-time energy density $\rhobar_\fd$ is given approximately as 
\beq
            \rhobar_\fd (t) ~\approx~ 50 \, \langle\phi\rangle^2 \, t^{-2}~.
\eeq 
In general, we see from Fig.~\ref{fig:fdabs} that our late-time energy density $\rhobar_\fd$
is independent of $\delta_G$ for \il{\lbrace \mbar t_G\lsim 10^3, \delta_G\lsim 0.5\rbrace}
and moreover scales roughly as 
\beq
      \rhobar ~\sim~ (\mbar t_G)^2
\label{4Dscaling}
\eeq
for \il{\mbar t_G \gsim 1}.
Of course, this scaling behavior extends to larger and larger values 
of $\mbar t_G$ as \il{\delta_G\to 0}, as we expect from the abrupt approximation.

\FloatBarrier
\section{\texorpdfstring{$N>1$}{N>1}:  ~A General Study with Multiple Fields\label{sec:FiniteModeNumber}}

We now consider how these results evolve as we move away from the \il{N=1} limit
and consider larger, arbitrary values of $N$.
Note that although we will eventually be considering the \il{N\to\infty} limit, the theories we analyze
here with finite \il{N>1} are interesting in their own right, as will be discussed further in the Conclusions.
As discussed in Sect.~\ref{sec:TheModel}, taking \il{N>1} introduces not only additional $\phi$ fields
(each potentially with its own critical oscillation time $t_\zeta$)
but also the non-trivial {\it mixing}\/ between these fields that is a consequence of the mass-generating
phase transition on the brane (resulting in a non-diagonal squared-mass matrix $\calM_{k\ell}^2$).
Systems with \il{N>1} can therefore be expected to be considerably more complex than their simpler
\il{N=1} cousins.

Towards this end, we shall concentrate not only on the behavior of the {\it total}\/ late-time energy density $\rhobar$
but also on its {\it distribution}\/ across the different $\phi$ fields.  As discussed in  
Sect.~\ref{sec:TheModel}, one cannot resolve the individual contributions to the total energy density $\rho$
during times when the mass matrix is appreciably changing, since the non-vanishing time-derivatives of 
$\calM_{k\ell}^2$ introduce new terms into $\rho$ which prevent its simple decomposition into the form in
Eq.~(\ref{energydecomp}).
However, at late times $t$ for which \il{(t-t_G)/\Delta_G \gg 1} ---
\ie, at times when the phase transition is largely completed --- 
the corresponding mass matrix $\calM_{k\ell}^2$ becomes essentially time-independent.
The mass-eigenstate fields $\phi_\lambda$ then decouple from each other and virialize.
At such late times, a decomposition
such as that in Eq.~(\ref{energydecomp})
becomes possible, with each such late-time energy density contribution $\rho_\lambda$ given in Eq.~(\ref{rholambda}).
Thus, in order to characterize the late-time energy configuration of 
the system with $N$ fields, we can calculate not only 
$\rhobar$ but also the individual contributions $\rhobar_\lambda$ corresponding to the
$N$ individual mass eigenstates $\phi_\lambda$.

Finally, following Ref.~\cite{DDM1}, another useful quantity we may define
is the so-called {\it tower fraction} $\eta$.   This quantity measures the 
fraction of the total energy density which is carried by all but the most abundant mode in the tower:
\beq\label{eq:etadefinition}
\etabar ~\equiv~ 1 - \max\limits_{\lambda}\left\{\frac{\rhobar_{\lambda}}{\rhobar}\right\} \ .
\eeq
For large but finite $N$, the tower fraction $\eta$ takes values within \il{0 \leq \eta < 1}
and can be viewed as quantifying the extent to which our system really has multiple components, with each carrying
some relevant portion of the total energy density.   If $\eta$ is extremely close to zero, 
then almost all of the total energy density is captured within a single field, rendering the other fields 
largely irrelevant from a phenomenological point of view.    For this reason we shall be most interested in systems with larger values of $\eta$. 
  
Given this, we shall analyze the general-$N$ system by tracing its energy flow, moment by moment through the mass-generating 
phase transition, ultimately evaluating the total late-time energy density
$\rhobar$, its distribution across the individual contributions $\rhobar_\lambda$, and the corresponding late-time value of $\etabar$ --- 
all as functions of $m$, $M_c$, $t_G$, and $\delta_G$.

\subsection{Instantaneous phase transition\label{instantsection}}

\begin{figure*}[t]
\includegraphics[keepaspectratio, width=0.9\textwidth]{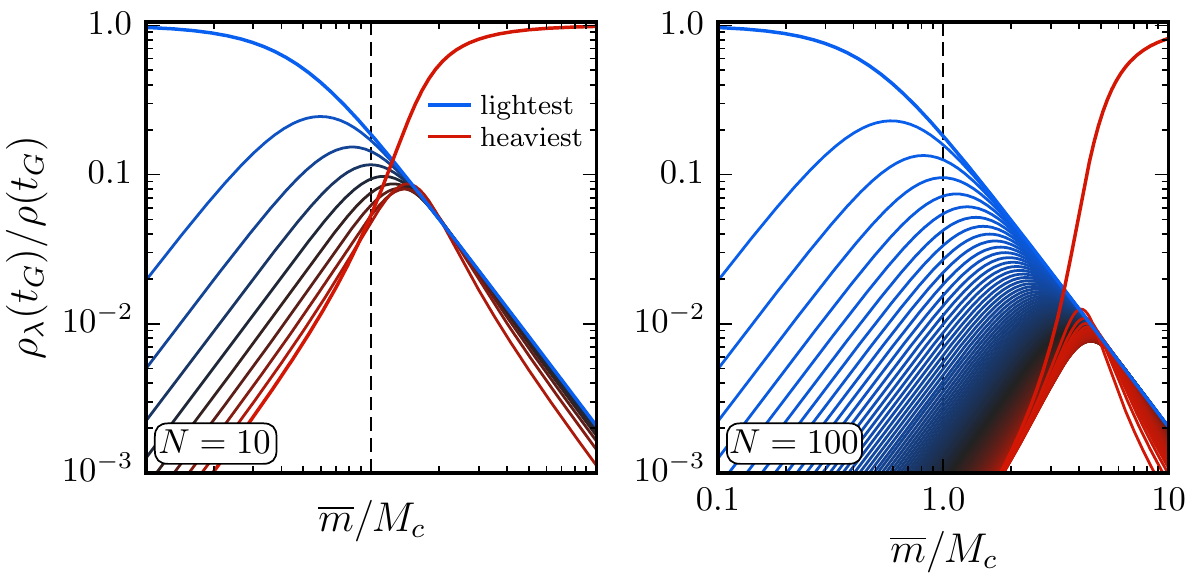}
\caption{The abundance fractions $\rho_\lambda/\rho$ at time \il{t=t_G}, evaluated within 
the abrupt \il{\delta_G\rightarrow 0}
limit and plotted as functions of $\mbar/M_c$ for \il{N=10} (left panel) and \il{N=100} (right panel).
In both panels we observe that the lightest mode carries the largest fractional abundance for small
$m/M_c$, but that this abundance is more fully distributed amongst the modes as
$m/M_c$ increases.   For finite $N$, this abundance then ultimately collects in the {\it heaviest}\/
mode as \il{m/M_c\to\infty}.    This last feature is ultimately an artifact of the truncation to only
finitely many modes, and disappears in the \il{N\to\infty} limit.   In all cases, 
the sum of the fractional abundances shown is equal to $1$ for all $m/M_c$.} 
\label{fig:initialabruptabundance}
\end{figure*}

It turns out that we can actually perform this calculation analytically in the special abrupt-transition 
limit \il{\delta_G\to 0}.
Indeed, in this limit our mass matrix $\calM_{k\ell}^2$ is time-independent both before and after $t_G$, and all that 
we need calculate is the sudden change of 
basis 
from the KK-eigenstate basis $\phi_k$ which is appropriate before $t_G$
to the mass-eigenstate basis $\phi_\lambda$ which is appropriate after.
Prior to $t_G$, only $\phi_{k=0}$ has an initial displacement \il{\langle \phi_0\rangle}: 
this field is massless prior to $t_G$, and thus the system has no energy and none of the fields oscillate.
However, for times after $t_G$, our mass eigenstates $\phi_\lambda$ become
\beq
          \phi_\lambda ~=~ \sum_{k=0}^{N-1} U_{\lambda k}\, \phi_k~
\label{basischange}
\eeq
where $U_{\lambda k}$ is the time-independent but $N$-dependent 
unitary basis-change matrix that diagonalizes $\calM_{k\ell}^2$.
The initial conditions in Eq.~(\ref{eq:initialconditions})
then become
\beqn
        && \phi_\lambda(t_G)  ~=~  U_{\lambda 0} \, \langle \phi_0\rangle~ \nonumber\\  
        && \dot \phi_\lambda(t_G) ~=~ 0~, 
\eeqn 
whereupon our total energy density at \il{t=t_G} is given by  
\beqn
      \rho(t_G) &=&  \sum_\lambda \rho_\lambda (t_G) = \half \sum_\lambda \lambda^2 \phi_\lambda^2 
                 \nonumber\\
             &=& \half \, \sum_\lambda \, \lambda^2 \, U_{\lambda 0}^2 \, \langle \phi_0\rangle^2 ~.
\eeqn
However, for all $N$,  our basis-change matrix $U_{\lambda k}$ satisfies the remarkable identity
\beq  
              \sum_\lambda \, \lambda^2 \, U_{\lambda 0}^2 ~=~ m^2 ~
\label{magic}
\eeq
which completely eliminates from $\rho(t_G)$ what has otherwise been a highly non-trivial 
dependence on $M_c$ (and $N$).
We thus find that our phase transition at \il{t=t_G} injects 
into our $N$-mode system 
a total energy density  
\beq
            \rho(t_G) ~=~ \half \, m^2 \, \langle \phi_0\rangle^2~, 
\label{rhotott}
\eeq
with individual fractional contributions
\beq
           {\rho_\lambda(t_G)\over \rho(t_G)} ~=~ \left( \lambda\over m\right)^2 U_{\lambda 0}^2~.
\label{rhoconts}
\eeq
Note that because we are working in the abrupt \il{\delta_G\to 0} limit,
the mass $m$, eigenvalues $\lambda$, and basis-change matrix $U_{\lambda k}$
in Eqs.~(\ref{basischange}) through (\ref{rhoconts})
are constant for all times after $t_G$ and thus equal to their late-time values:
\il{m=\mbar}, \il{\lambda=\lambdabar}, and
\il{U_{\lambda k} = \overline U_{\lambda k}}.

In general, we see from Eq.~(\ref{rhotott}) that the total energy density
injected into our system grows polynomially with $m$.
However, the individual contributions to this total energy density 
vary non-trivially as a function of $m/M_c$ --- even when expressed 
as a fraction relative to the total.
These individual fractional contributions to the total energy density 
at \il{t=t_G}
are plotted as functions of $m/M_c$ 
for \il{N=10} in the left panel of Fig.~\ref{fig:initialabruptabundance} 
and
for \il{N=100} in the right panel of Fig.~\ref{fig:initialabruptabundance}. 
In each case, we see that these energy-density distributions
are highly sensitive to $m/M_c$,
with the majority of the energy density at \il{t=t_G}
carried by the lightest mode for small \il{m/M_c}
(as consistent with our expectation from the 4D limit).
However, as $m/M_c$ increases,
this picture changes, with more and more of the states beginning to carry
greater fractions of the total energy density.
Note that for all values of $m/M_c$, the sum of these fractional
energy densities is fixed at $1$.

Ultimately, for very large \il{m^2/M_c^2\gsim {\cal O}(\sqrt{N})},
the majority of the energy density 
begins to collect in the {\it heaviest}\/ mode.
However, we have already seen in Sect.~\ref{sec:TheModel}
that this is precisely the regime in which the heaviest eigenvalue $\lambda_{N-1}$ 
begins to diverge.
Indeed, as discussed in Sect.~\ref{sec:TheModel}, 
this feature is ultimately an artifact of truncating our system to a finite
number of modes, with \il{N<\infty}.
A similar situation persists for the corresponding energy densities.
Even though the energy density associated with the heaviest mode tends to dominate
in the region with \il{m^2/M_c^2\gsim {\cal O}(\sqrt{N})},
this region becomes increasingly remote
as \il{N\to \infty} and effectively vanishes.
Thus, for \il{m/M_c\gg 1} and \il{N\to\infty}, all of the modes in our theory 
tend to share the total energy density
essentially equally.

\begin{figure}[t]
\includegraphics[keepaspectratio, width=0.5\textwidth]{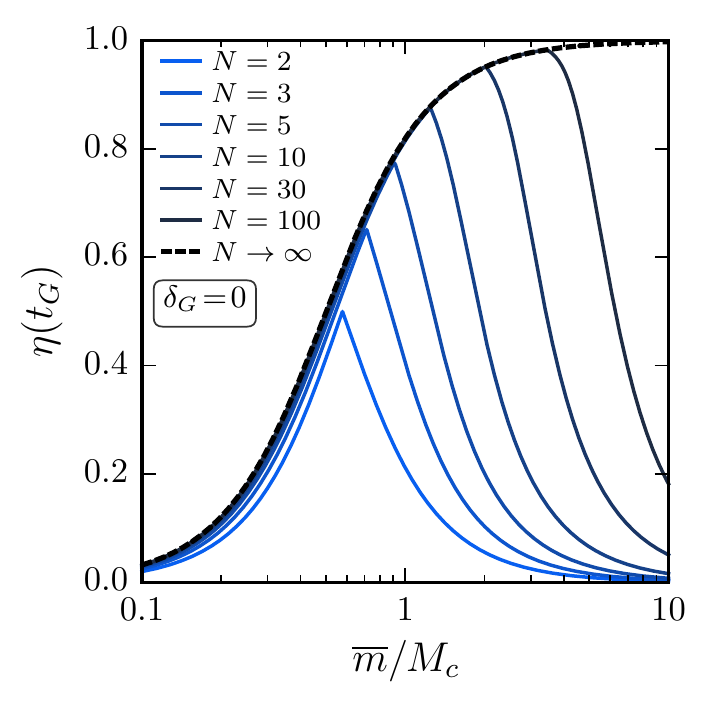}
\caption{The tower fractions $\eta(t_G)$ corresponding to the fractional
abundances shown in Fig.~\ref{fig:initialabruptabundance},
plotted as functions of $m/M_c$ for different values of $N$.
We see that the tower fraction begins near zero for small $m/M_c$ but then begins to grow as
$m/M_c$ increases.    For finite $N$, the tower fraction ultimately hits a peak before returning
to zero at large $m/M_c$, while in 
the \il{N\to\infty}  limit the tower fraction continues monotonically to $1$.}
\label{fig:towerfraction}
\end{figure}

In Fig.~\ref{fig:towerfraction}
we show the
corresponding values of the tower fraction $\eta(t)$ at \il{t=t_G}, plotted 
as functions of $m/M_c$.   
As we see, $\eta(t_G)$ is non-monotonic as a function of
$m/M_c$ for all finite values of $N$, starting near \il{\eta\approx 0} for small
$m/M_c$ and reaching a maximum at a certain critical
$N$-dependent value $(m/M_c)_{\rm peak}$  
before declining back to \il{\eta\approx 0} for large $m/M_c$.
This non-monotonic behavior for finite $N$ is ultimately a result of the fact
that the identity of the mode carrying the maximum abundance is itself
a function of $m/M_c$, with the lightest mode carrying the largest
abundance for small $m/M_c$ but the heaviest mode carrying the largest
abundance for large $m/M_c$.   However, as \il{N\to\infty}, we find
that the magnitude of the peak tends towards unity while
the position of peak itself shifts towards increasingly large values of $m/M_c$.
Indeed, with excellent precision, we find that the position of the peak is given by
\beq
    \left( {m\over M_c}\right)_{\rm peak} \approx~ {4\over 9} \, N^{4/9}~,
\eeq
whereupon it follows that \il{(m/M_c)_{\rm peak}\to\infty} as \il{N\to \infty}.
We thus find that $\eta(t_G)$ becomes completely monotonic in the infinite-$N$ 
limit, transitioning from \il{\eta\approx 0} at small $m/M_c$ to \il{\eta \approx 1} at large $m/M_c$.

The results illustrated in Fig.~\ref{fig:initialabruptabundance} and \ref{fig:towerfraction}
apply to the energy configuration 
of our system at \il{t=t_G}.
A natural question, however, concerns the extent to which these results apply for the {\it late-time}\/
energy-density fractions $\rhobar_\lambda/\rhobar$ and late-time tower fraction $\etabar$. 
Even though the mass matrix $M_{k\ell}^2$ and  associated eigenvalues 
are time-independent for all \il{t> t_G} in the abrupt \il{\delta_G\to 0} limit,   
the mass eigenstates $\phi_\lambda(t)$ 
and corresponding energy densities $\rho_\lambda(t)$ nevertheless 
continue have a non-trivial time-dependence.    In general, 
a single field $\phi_\lambda$ with constant mass $\lambda$
remains essentially fixed for all \il{t < t_\zeta^{(\lambda)} \equiv\kappa/(2\lambda)}
and then transitions to damped-oscillatory behavior for \il{t> t_\zeta^{(\lambda)}}.
Likewise, the corresponding energy density is essentially fixed for  
\il{t < t_\zeta^{(\lambda)}}
and then decays as $t^{-\kappa}$ for 
\il{t > t_\zeta^{(\lambda)}}.
(Exact analytical results for the fields and energy densities can be found
in Appendix~A and Fig.~1 of Ref.~\cite{TwoTimescales}.)
Thus, in our present situation with an abrupt phase transition at \il{t=t_G},
all of our fields 
will already be underdamped and commence oscillations 
simultaneously 
at $t_G$
as long as \il{t_\zeta^{(\lambda)} \leq t_G} for all $\lambda$. 
In such cases, the ratios $\rho_\lambda/\rho$ as well as the corresponding
tower fraction $\eta$ will be essentially time-independent, with the same values
at late times as they have at $t_G$.

\subsection{Phase transition with arbitrary width \texorpdfstring{$\delta_G$}{delta_G}}

The results given above apply only in the abrupt \il{\delta_G\to 0} limit.
However, 
we now wish to explore the full parameter space  and calculate our total and fractional
late-time energy  densities and tower fractions throughout the 
\il{(\mbar t_G, \delta_G)} plane.
We already did this for the \il{N=1} limiting case in Sect.~\ref{sec:FourDimensionalLimit},
where we found that the effects of taking \il{\delta_G>0} tended to produce either
suppressions or enhancements in the resulting late-time energy \il{\rhobar_{\rm 4D}(\delta_G)} compared
with $\rho_\fd(0)$  depending on the precise values of $\mbar t_G$ and $\delta_G$.
These results were shown in the left panel of Fig.~\ref{fig:singlefielddensityplot}.
We therefore now seek to know what additional effects 
beyond those in Fig.~\ref{fig:singlefielddensityplot}
emerge from considering arbitrary values of $N$ rather than merely \il{N=1}.

\begin{figure*}[bht]
\includegraphics[keepaspectratio, width=0.9\textwidth]{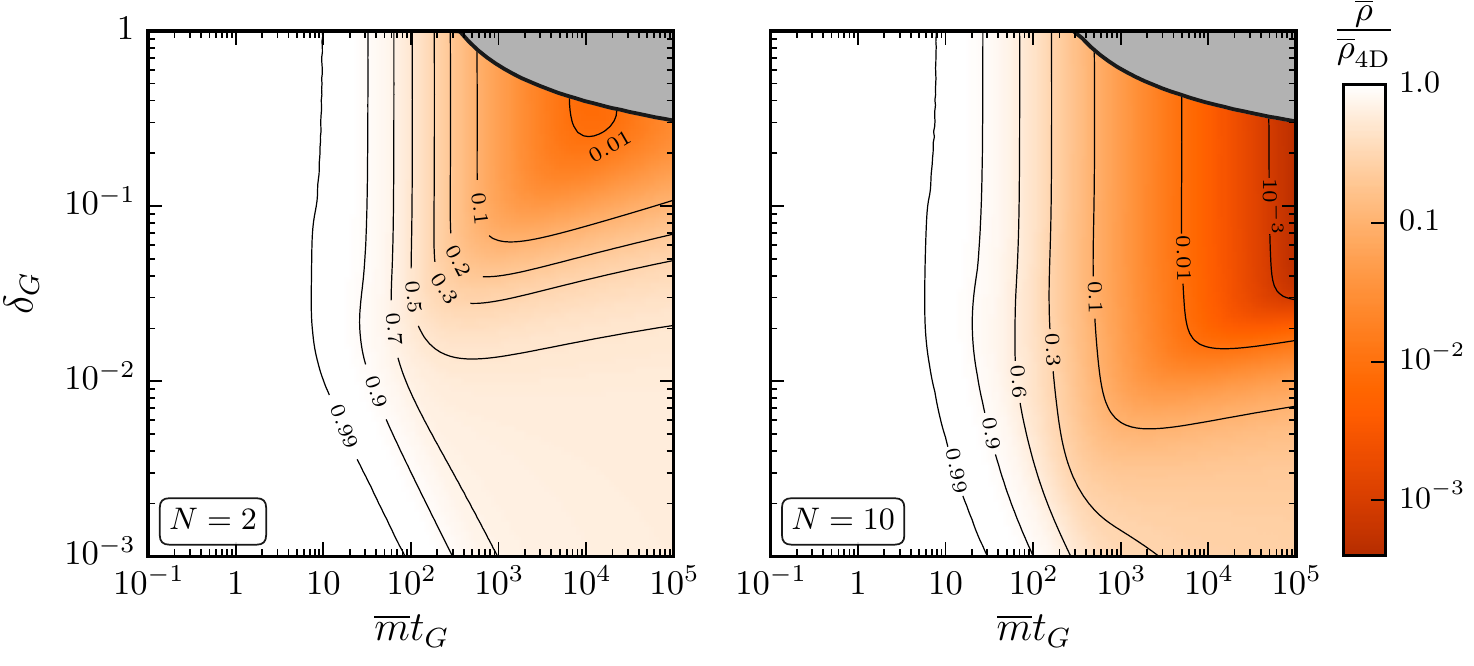}
\caption{The total late-time abundance $\rhobar$ for $N$ modes, 
evaluated within the \il{(\mbar t_G,\delta_G)} plane
for \il{N=2} (left panel) and \il{N=10} (right panel)
and expressed as a fraction of the 
corresponding value $\rhobar_\fd$ for the 4D \il{N=1} special case.
We have chosen \il{t_G=10^2/M_c} as a reference value for both plots above.
We see that the introduction of additional modes relative to the 4D special case
either {\it preserves}\/ the 4D result nearly exactly for \il{\mbar t_G\lsim 10}  or
{\it suppresses}\/ it for \il{\mbar t_G\gsim 10}, with this suppression becoming
increasingly severe for larger $\mbar t_G$ (or equivalently, larger $\mbar/M_c$)
and \il{\delta_G\lsim 1}.  Note that the gray regions in the upper right corners of
these panels (and in similar subsequent figures throughout this paper) 
are excluded for the reasons discussed in the paragraph below Eq.~(\ref{bigwidth}).  }
\label{fig:finitemodedensityplotsgenN}
\end{figure*}

Our results are shown in Fig.~\ref{fig:finitemodedensityplotsgenN} for
\il{t_G=10^2/M_c}.
Remarkably, 
for that portion of the \il{(\mbar t_G,\delta_G)} parameter space
  with \il{\mbar t_G\lsim 10}, 
{\it we find that the total late-time energy density $\rhobar$
is virtually identical to what it would have been for only one mode!}\/ 
Of course, this behavior is consistent with what we already saw for the 
abrupt \il{\delta_G\to 0} limit and very small $m/M_c$, but what is remarkable
is that this behavior extends even for non-zero $\delta_G$ and larger $m/M_c$
as well.
Moreover, for \il{\mbar t_G\gsim 10}, 
we find that our
late-time energy density is actually
suppressed rather than enhanced
relative to our 4D expectations --- 
all this despite the presence of extra modes whose masses are also lifted by 
(and which therefore also receive an additional energy-density insertion from)
the mass-generating phase transition.

These results may be understood as follows.   
Since we have taken \il{t_G= 10^2/M_c} as a reference value,
we see that \il{\mbar t_G = 10^2 \, \mbar/M_c}.
Thus values \il{\mbar t_G \lsim 10^2} correspond to 
\il{\mbar/M_c\lsim 1}, and we have already seen that 
the bulk of the energy density remains concentrated  
in the lightest mode for such values of $\mbar/M_c$.
As a result, 
our system is functionally no different than the single-mode 4D system
for small values of $\mbar t_G$. 
By contrast, as we increase the value of $\mbar t_G$, the total energy density 
of our system is more equally distributed across the different modes.
The oscillations of the heavier modes then dissipate the
energy density more rapidly than the lighter modes during the phase transition, resulting in a more
rapid dissipation of the total energy density and thus an overall suppression 
of the late-time energy density as compared with 4D expectations.
It also is important to note that this latter effect requires \il{\delta_G>0}, even
for large $\mbar t_G$.
Indeed, if \il{\delta_G=0}, our energy density is partitioned across the different
modes precisely as described in Sect.~\ref{instantsection},
whereupon the unitarity relation in Eq.~(\ref{magic})
removes all dependence on $N$ and ensures the same results as we would
have had in 4D!

\begin{figure*}[t]
\includegraphics[keepaspectratio, width=0.9\textwidth]{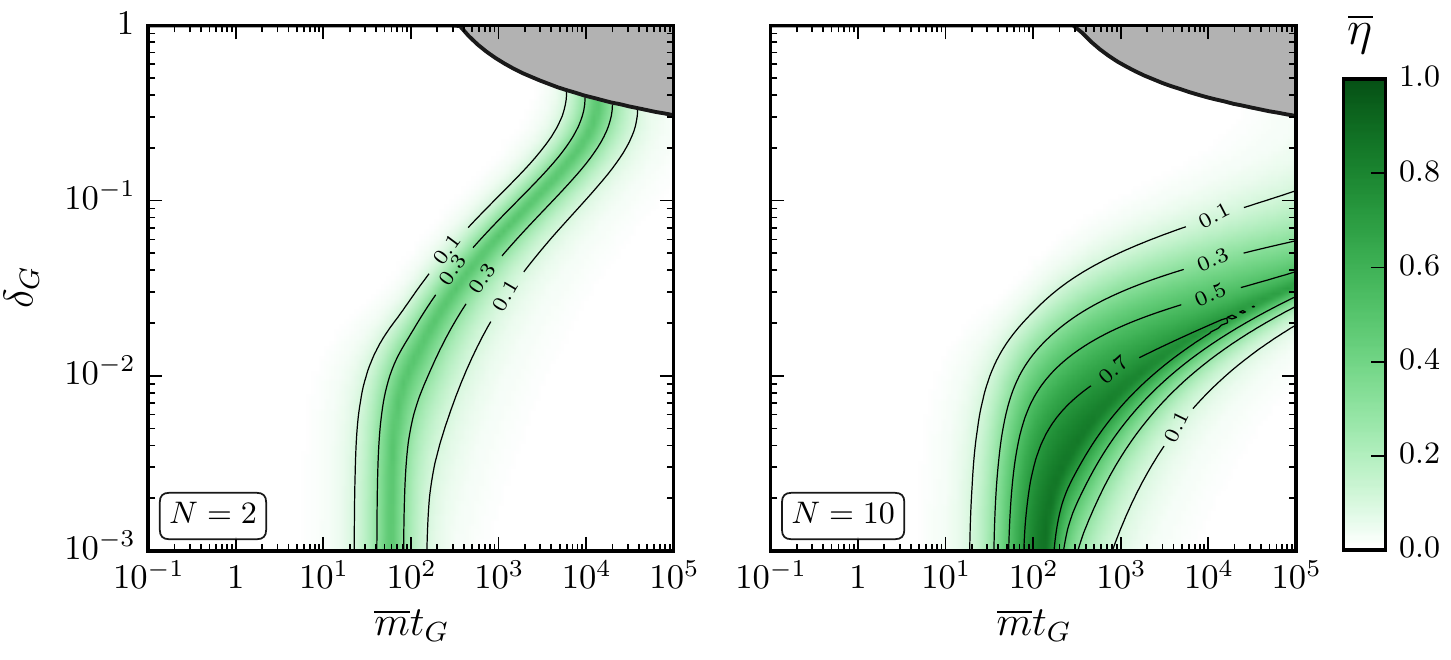}
\caption{The total late-time tower fraction $\etabar$ corresponding to the panels
shown in Fig.~\protect\ref{fig:finitemodedensityplotsgenN}.
In each case we see that the \il{(\mbar t_G,\delta_G)} plane is subdivided into
two disjoint \il{\etabar\approx 0} regions (white) by a narrow \il{\etabar>0} ribbon (green):    
the lightest mode carries the bulk of the energy density in the region to the left of the ribbon,
while the heaviest mode carries the bulk of the energy density in the region to the right.
Interestingly, the division between these two regions 
is non-monotonic as a function of not only $\mbar t_G$ but also $\delta_G$.}
\label{fig:finitemodedensityplotsgenN2}
\end{figure*}

In Fig.~\ref{fig:finitemodedensityplotsgenN2} we plot the corresponding values of the
late-time tower fraction $\etabar$.
As evident from this figure, the \il{(\mbar t_G, \delta_G)} plane is subdivided
into two disjoint \il{\etabar \approx 0} regions by 
a fairly narrow ``mountain range'' along which \il{\etabar>0}.
As we already determined in Sect.~\ref{instantsection} for the \il{\delta_G\to 0} limit,
the region with smaller $\mbar t_G$ (or equivalently smaller $\mbar/M_c$) 
corresponds to the region in which the lightest mode carries all of the energy density.
By contrast, as we dial $\mbar t_G$ towards greater values, the energy density
is increasingly distributed  
across all the modes, resulting in a greater tower fraction. 
Ultimately, however, as we dial \il{\mbar t_G \to\infty} in our finite-$N$ system,
all of the energy density finds itself in the heaviest mode, and the tower fraction returns
to zero.  This behavior is consistent with that shown in 
Fig.~\ref{fig:towerfraction} for the \il{\delta_G\to 0} limit.

This explains the non-monotonicity of the late-time tower fraction $\etabar$ 
as a function of $\mbar t_G$.
However, we additionally learn from Fig.~\ref{fig:finitemodedensityplotsgenN2} that 
the late-time tower fraction is also non-monotonic as a function of $\delta_G$.
This too is relatively straightforward to explain.
For $\mbar t_G$ sufficiently large and for small $\delta_G$, we have already seen that the bulk of the
energy density is carried by the heaviest mode. 
However, as we increase $\delta_G$, we are increasing the timescale associated with our phase 
transition --- \ie, the timescale over which energy is pumped into the system
and each mode becomes populated.    
Eventually this timescale reaches the timescale associated with the oscillations of the heaviest mode.
This then suppresses the energy transfer into the (nevertheless dominant) heaviest mode, 
thereby increasing the tower fraction. 
Indeed, as we increase $\delta_G$ still further, we begin to suppress the energy densities of lighter
and lighter modes, and this too has the effect of increasing the tower fraction.
Eventually, however,
as we continue to increase $\delta_G$, 
we reach a point
where we have suppressed the majority of the modes and the bulk of the energy density
begins collecting in the lightest mode.   Subsequent increases in $\delta_G$ then only
reinforce the dominance of the lightest mode, thereby causing 
the tower fraction to return to zero.
Indeed, as shown in Fig.~\ref{fig:finitemodedensityplotsgenN2}, we have crossed the green 
``mountain range'', leaving us
in the region in which the lightest mode carries all of the energy density.

Thus far, our discussion has focused on the regime with \il{t_G M_c \gg 1}.
In this regime, all of our excited modes are underdamped and have begun oscillating at very early 
times prior to the phase transition.   This will be true even for the lightest mode
if $\mbar t_G$ is chosen sufficiently large.
However, if \il{t_G M_c \ll 1} and \il{N\gsim 1/(t_G M_c)},  
only an upper subset of the modes will have begun oscillating
prior to the phase transition --- 
the lightest modes will remain overdamped and fixed.
Thus, the introduction of the heavier modes relative to the 4D case has the 
effect of distributing some of the energy density into heavier modes
which begin oscillating earlier than the lightest mode, 
thereby enhancing the dissipation of the total energy
density relative to the 4D case. 
It is important to stress that this source of suppression is
completely distinct from that discussed above, resulting instead from
differences in the times at which individual modes begin oscillating.
As such, this effect --- which was originally discussed in Ref.~\cite{DDGAxions}
within the context of the 
abrupt \il{\delta_G\to 0} limit --- is largely independent of $\delta_G$.

\FloatBarrier
\section{Approaching Asymptotia: ~\texorpdfstring{\il{N\to \infty}}{N->infty}\label{Approaching}}


\begin{figure*}
\includegraphics[keepaspectratio, width=1.0\textwidth]{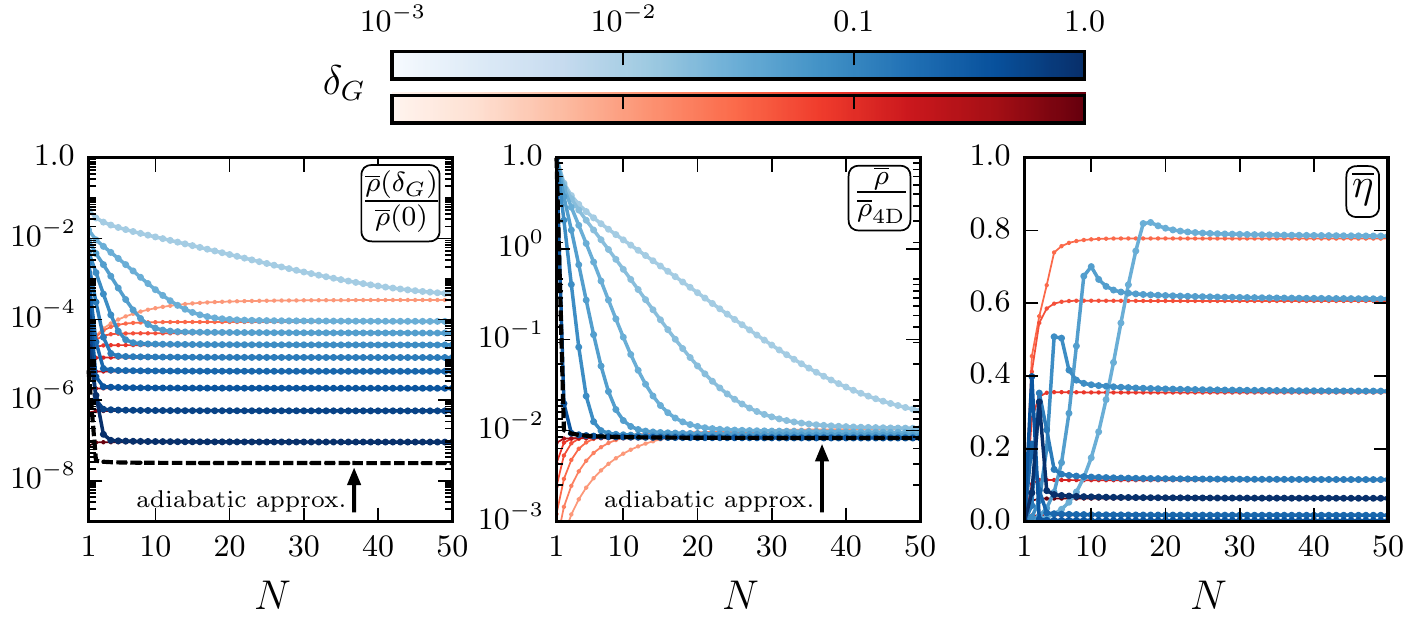}
\caption{
The total late-time abundance $\rhobar$ and late-time tower fraction $\etabar$ of our $N$-mode system,
plotted as functions of $N$ for different values of $\delta_G$. The left and center panels show 
the total late-time abundance expressed as a fraction of the value it would have in the abrupt 
\il{\delta_G\to 0} and 4D limits, respectively, while the right panel shows the late-time tower fraction.
For all three panels we have taken  \il{t_G=10^2/M_c} and \il{\mbar=10^2 M_c}, thereby fixing 
\il{\mbar t_G=10^4} as a benchmark value. The blue curves are calculated using the usual ``IR-based'' 
truncation which has been employed thus far in Sects.~\ref{sec:TheModel} through \ref{sec:FiniteModeNumber},
while the red curves are calculated using the alternative ``UV-based'' truncation to be discussed in 
Sect.~\protect\ref{alttrunc}; both truncations have the same \il{N\to\infty} asymptotic behavior
but the UV-based truncation reaches asymptotia far more rapidly and smoothly. The black-dashed line that 
occurs in each of the energy-density panels indicates the results of the multi-component adiabatic 
approximation of Eq.~\eqref{eq:multiadiabaticapproximation} which serves as a lower bound for the energy 
density of the ensemble.}
\label{fig:Nscaling}
\end{figure*}

We now wish to study how our $N$-mode system evolves as a function of $N$, 
with an eye towards understanding the asymptotic behavior of our system for large \il{N\gg 1}.
This will ultimately enable us to extract the behavior of the 
full \il{N=\infty} Kaluza-Klein tower, as we shall do in Sect.~\ref{sec:KKLimit}.~ 
This will also enable us to understand the effects that come from truncating
the KK system to finite but large $N$. 

\subsection{Large-\texorpdfstring{$N$}{N} behavior:  The road to asymptotia\label{road}}

In Fig.~\ref{fig:Nscaling}, the blue curves indicate the $N$-dependence of our 
two quantities of interest, namely
the late-time energy density $\rhobar$ of our $N$-mode system 
as well as its corresponding late-time tower fraction $\etabar$.
While the latter quantity is shown in the right panel,
the former is plotted in the left panel 
as a fraction of the value it would
have in the abrupt \il{\delta_G\to 0} limit 
and in the middle panel 
as a fraction of its corresponding 4D value.
For all three panels we have taken  \il{t_G=10^2/M_c}
and \il{\mbar=10^2 M_c}, thereby fixing \il{\mbar t_G =10^4} as a benchmark value.
We have then 
plotted the resulting curves 
(blue) as functions of $N$ 
for different values of $\delta_G$.
(By contrast, the red curves will be discussed in Sect.~\ref{alttrunc}.)

As expected, each of the blue curves shown in Fig.~\ref{fig:Nscaling} eventually
heads towards a finite asymptote as \il{N\to\infty}.
One immediate observation from the left and middle panels of 
Fig.~\ref{fig:Nscaling}
is that increasing the number of modes
in our system 
generally results in an {\it increasing suppression}
of the total late-time energy density when compared to the
values this energy density would have in the abrupt \il{\delta_G\to 0} or
4D limits.
Thus, {\it the larger the value of 
$N$, the more the abrupt and 4D approximations fail to accurately
estimate the late-time energy density}\/.
This is an important result, given that most approaches to calculating
the energy densities of such multiple-component systems in the literature
assume that the phase transition occurs when the fields are still overdamped.
Compared with these static fields, any such phase transition is therefore essentially occurring
with an infinitely short timescale $\delta_G$, and is thereby functionally equivalent to the abrupt
approximation.    The results shown in Fig.~\ref{fig:Nscaling} 
thus illustrate what happens as one moves away from these assumptions. 

In this vein, it is perhaps also worthwhile to consider the adiabatic 
approximation which is traditionally applied in single-component
scalar theories undergoing mass-generating phase transitions.  
Of course, while an adiabatic approximation can be realized
for a single mode, we do not expect such an approximation to be appropriate
for a large collection of modes with vastly different masses and non-trivial mixings.
Despite this, we can nevertheless imagine limiting cases in which
all excited modes above the lightest mode are
drained of significant energy density.    
In such a regime we can then imagine applying the adiabatic approximation 
to the only surviving mode $\phi_{\lambda_0}$ in the ensemble, leading to a
definition
\beq
\rho(t)|_{\rm ad}  ~\equiv ~ \frac{1}{2}\langle\phi_0\rangle^2 \lambda_0(t^{(\lambda_0)}_{\zeta}) \, 
\lambda_0(t)\left[\frac{a(t^{(\lambda_0)}_{\zeta})}{a(t)}\right]^3 \ ,
\label{eq:multiadiabaticapproximation}
\eeq
where $t^{(\lambda)}_{\zeta}$ is defined as the time at which \il{3H(t) = 2\lambda(t)}
and where we implicitly assume \il{\delta_G=1} when evaluating $\lambda(t)$.
We have already seen in the left panel of Fig.~\ref{fig:singlefielddensityplot} that the
adiabatic approximation sets a lower bound for the energy density associated with
a single mode --- indeed, in all other regions of parameter space the energy density 
is either the same or enhanced.
Likewise, the approximation in Eq.~(\ref{eq:multiadiabaticapproximation})
consists of disregarding whatever energy density might reside in the excited modes.
Thus, we expect the adiabatic approximation in 
Eq.~(\ref{eq:multiadiabaticapproximation})
to provide us with a lower bound on the total energy density of our $N$-mode
system throughout all regions of parameter space.

The dashed-black curves in the left and middle panels of Fig.~\ref{fig:Nscaling} show the 
behavior of Eq.~\eqref{eq:multiadiabaticapproximation} as a function of $N$. 
Indeed, as expected, we see that \emph{the adiabatic approximation serves 
as a lower bound on the total energy density of the tower}, just as it does in the single-field
scenario.  Indeed, we see from Fig.~\ref{fig:Nscaling} 
that the adiabatic limit seems to be reached most directly for large $N$
(such as for a full KK tower) and \il{\delta_G\to 1}.
By contrast, the results in other regions of parameter space differ significantly from this limit.

\subsection{A new truncation of KK theory:  An alternate road to asymptotia\label{alttrunc}}

Our results in Fig.~\ref{fig:Nscaling} clearly illustrate a successful road to asymptotia,
as each of the blue curves heads towards a finite asymptotic value as \il{N\to \infty}.
This is not a surprise, since
we began our analysis by truncating our infinite-dimensional KK mass matrix
$\calM_{k\ell}^2$ in Eq.~(\ref{kmassmatrix}) by 
retaining only its first $N$ rows and columns.
Thus it is natural that taking $N\to\infty$ restores the physics of the infinite-matrix limit.
Indeed, truncating our mass matrix in
Eq.~(\ref{kmassmatrix}) 
to only its first $N$ rows and columns
represents one way of constructing a finite $N$-mode theory 
whose \il{N\to\infty} limit reproduces the physics of the full KK tower 
at any moment in time.  

However, this is not a unique truncation to a finite $N$-mode theory.
There is, in fact, an alternative truncation 
of our full KK theory at any moment in time which also has only $N$ modes and which also reproduces the physics
of the full KK tower as \il{N\to \infty}, yet yields significantly different physical results for finite $N$.
Indeed, as we shall see, this alternate truncation exhibits an even more rapid path
to asymptotia, yielding what are essentially the full infinite-$N$ values of
the late-time energy densities and tower fraction for even smaller values of $N$
than are required using the standard truncation that we have employed thus far.

Recall that our standard truncation began with the full, infinite-dimensional
mass matrix $\calM^2_{k\ell}$ in Eq.~(\ref{kmassmatrix}), expressed in the KK basis of the individual KK modes $\phi_k$.
This matrix was then truncated to its first $N$ rows and columns.
The resulting \il{N\times N} matrix then defined our truncated KK theory,
and all subsequent calculations proceeded from this truncated matrix.
Specifically, all calculations were performed directly from 
the equations of motion~(\ref{eq:KKequationsofmotion})
in the KK basis (which has the advantage of representing a basis choice that does not change with time, even in the presence of the mass-generating phase transition), 
and the results at late times were 
converted to the mass-eigenstate basis (such as for quoting late-time quantities such as $\rhobar_\lambda$ or $\etabar$ which
pertain to individual mass eigenstates) only at the final step.

However, an alternative approach is {\it not}\/ to truncate our infinite-dimensional mass matrix 
in the KK basis $\phi_k$,  but rather to transform the infinite-dimensional matrix into its
mass-eigenstate basis and then truncate the resulting {\it mass-eigenstate}\/ matrix.
In general, this matrix will take the form
\beq
\mathcal{M}_{\lambda\lambda'}^2 ~=~ 
\begin{pmatrix}[1.1]
\lambda_0^2 & 0 & 0 & \cdots \\
0 & \lambda_1^2 & 0 & \cdots  \\
0 & 0 & \lambda_2^2 & \cdots  \\
\vdots & \vdots & \vdots & \ddots
\end{pmatrix} \ ,
\label{eq:massbasismassmatrix}
\eeq
where the $\lambda_k^2$ are the mass eigenvalues that are calculated in 
the infinite-$N$ limit at any moment in time.   These eigenvalues will be discussed below.

Note that truncating the mass matrix in the mass-eigenstate basis 
is mathematically different than truncating the mass matrix in the KK-eigenstate basis.
Of course, these procedures are in some sense parallel in that they both
begin from equivalent infinite-dimensional matrices which describe the 
same KK system and which are related to each other through a simple,
unitary basis change.   
However, their truncations to \il{N\times N} submatrices are mathematically different, and thus have
different physical effects.

Having truncated our mass-eigenstate matrix $\calM^2_{\lambda \lambda'}$
in Eq.~(\ref{eq:massbasismassmatrix}),
our final step is to convert the resulting
matrix back to the KK basis.
In order to do this, we 
use a similarly truncated version of the exact basis-change matrix $U_{\lambda k}$ 
that
would have related $\calM^2_{k\ell}$ and $\calM^2_{\lambda\lambda'}$ in the full infinite-dimensional 
limit.
Specifically, we define
\beq
     \widetilde \calM^2_{k\ell} ~\equiv ~ \sum_{\lambda,\lambda'} \, 
              (\widehat U^\dagger)_{k\lambda} \, \widehat \calM^2_{\lambda\lambda'}  \, \widehat U_{\lambda' \ell} 
\label{tildeM}
\eeq
where $\widehat\calM^2_{\lambda\lambda'}$ and $\widehat U_{\lambda'\ell}$ here represent the first 
\il{N\times N} submatrices within the exact infinite-$N$ matrices 
$\calM^2_{\lambda\lambda'}$ and $U_{\lambda k}$ respectively.
Note that since $\widehat U$ is a truncated version of the unitary infinite-$N$ matrix $U$,
the truncated matrix $\widehat U$ is not unitary by itself.   In particular, $\widehat U$ is {\it not}\/ the matrix 
that diagonalizes the new matrix $\widetilde \calM^2_{k\ell}$ defined in Eq.~(\ref{tildeM}), and thus, strictly speaking, 
we should not regard $\widetilde \calM^2_{k\ell}$
defined in Eq.~(\ref{tildeM})
as representing the KK-basis version of $\widehat \calM^2_{\lambda\lambda'}$.
We can nevertheless proceed to use $\widetilde\calM^2_{k\ell}$ directly in our equations of motion~(\ref{eq:KKequationsofmotion}), 
treating $\widetilde\calM^2_{k\ell}$ as we would any other mass matrix.
Converting our final results 
into statements about individual mass eigenstates 
at late times is then done in the usual way by 
diagonalizing $\widetilde\calM^2_{k\ell}$ and calculating its mass eigenvalues and eigenvectors.
Note that these eigenvalues will generally differ from the $\lambda_i$ which
appear in the 
$\widetilde\calM^2_{k\ell}$ matrix
and only approach these $\lambda_i$ as \il{N\to \infty}.

Because this alternate truncation of our KK theory
utilizes the exact eigenvalues $\lambda_k$  and exact basis-change matrix elements $U_{\lambda k}$ corresponding
the full KK theory, it may at first glance seem that this    
alternate truncation cannot be realized in practice.
However, it turns out to be relatively straightforward to solve the eigensystem corresponding
to the full infinite-dimensional mass matrix $\calM^2_{k\ell}$ in Eq.~(\ref{kmassmatrix}) --- not only 
numerically, but even analytically.
One finds~\cite{DDGAxions,DDM1} 
that the mass eigenvalues $\lambda_i$ at any moment in time are the (infinite) set
of solutions to the transcendental equation
\beq
         {\pi \lambda \over M_c} \, \cot\left( \pi \lambda\over M_c\right) ~=~ {\lambda^2\over m^2}~,
\label{eigeq}
\eeq
 and likewise the corresponding $U_{\lambda k}$ matrix is given by
\beq
            U_{\lambda k} ~=~ \left(  {r_k \lambda^2 \over \lambda^2 - k^2 M_c^2} \right) \, A_\lambda
\eeq
where the $r_k$ are defined below Eq.~(\ref{eq:orbifoldmodeexpansion}) and
where 
\beq
           A_\lambda~\equiv  ~  {\sqrt{2} \, m^2 \over \lambda \sqrt{ \lambda^2 + m^2 + \pi^2 m^4/M_c^2 }}~.  
\label{Adefn}
\eeq
Following the procedure outlined above, we then find that our alternate mass matrix
is given by
\beq
          \widetilde \calM^2_{k\ell} ~=~  \sum_{\lambda=\lambda_0}^{\lambda_{N-1}}  { r_k r_l  \, A_\lambda^2  \, \lambda^6 \over
                  (\lambda^2 - k^2 M_c^2)
                  (\lambda^2 - \ell^2 M_c^2)}~.
\label{massmatrixnew}
\eeq
In this connection, we remark 
that although constructing $\widetilde \calM^2_{k\ell}$ requires  
explicit knowledge 
of the exact eigenvalues $\lambda_i$ and matrix elements $U_{\lambda k}$, 
this in no way implies that we 
have already solved the problems we originally set out to investigate.
Indeed, our goals are far deeper than mere KK spectroscopy and instead pertain  
to understanding the dynamical energy flow within our KK system in the presence of 
mode-mixing and mass-generating phase transitions.

Although each element of the matrix $\widetilde\calM^2_{k\ell}$ in Eq.~(\ref{massmatrixnew}) depends on $N$ 
explicitly through the upper limit of the $\lambda$-summation, 
this matrix smoothly reproduces our original $\calM^2_{k\ell}$ mass matrix  
as \il{N\to\infty}.
Indeed, through clever use of the eigenvalue equation~(\ref{eigeq}), 
cotangent summation identities such as  
\beq
      \sum_{k=0}^\infty {2 \lambda^2 \over \lambda^2-k^2 M_c^2} ~=~ 1 
         + {\pi \lambda \over M_c} \, \cot\left( \pi \lambda\over M_c\right)~, 
\eeq
and unitarity relations such as that in Eq.~(\ref{magic}),
it is possible to demonstrate explicitly that performing the summation in Eq.~(\ref{massmatrixnew})
over the {\it full}\/ spectrum (\ie, taking \il{N\to\infty}) 
reproduces our original mass matrix in Eq.~(\ref{kmassmatrix}), as it must by construction.
For example, in what is perhaps the simplest case, we see from Eq.~(\ref{massmatrixnew}) that
\beq
         \widetilde \calM^2_{00} ~=~   \sum_{\lambda=\lambda_0}^{\lambda_{N-1}}  A_\lambda \, \lambda^2~.
\eeq
If the summation had proceeded over the entire infinite spectrum, the unitarity relation in
Eq.~(\ref{magic}) would have given us the correct infinite-$N$ result \il{\calM^2_{00} = m^2}.
Thus, we see in this case 
that our UV-based truncation consists of gently draining away the contributions   
to the unitarity sum that come from the most massive modes, thereby deforming this
first matrix element in a gentle, $N$-dependent way.
Other matrix elements are similar.

The new matrix $\widetilde \calM^2_{k\ell}$ in Eq.~(\ref{massmatrixnew}) 
thus defines an alternate truncation of the full KK theory. 
Indeed, both our original truncated mass matrix $\calM^2_{k\ell}$ in Eq.~(\ref{kmassmatrix}) and our new mass matrix
$\widetilde \calM^2_{k\ell}$ 
in Eq.~(\ref{massmatrixnew})
describe the same KK theory in their \il{N\to\infty} limits.
However, for any finite $N$ these mass matrices define distinct theories.   
We shall refer to our traditional truncation as being ``IR-based'', since it builds our finite-$N$ theory
directly from the ground up, KK mode by KK mode, directly as they were in the full theory without regard for any 
of the UV physics.
By contrast, we shall refer to our alternative truncation as being ``UV-based'' in the sense that it utilizes
the full UV values of the mass eigenvalues and basis-changing matrices prior to truncation,
and gently builds these quantities into our recipe for truncation. 
This is thus more of a ``top-down'', UV-sensitive approach.
Of course, it is only because of the mixing of KK states induced by the phase transition on the brane 
that these two truncations are distinct.

What makes this alternate UV-based truncation 
particularly important for our purposes in this paper
is that it provides a much more rapid road to asymptotia than does our usual truncation.
In other words, the asymptotic values of late-time quantities such as $\rhobar_\lambda$
and $\etabar$ are approached more rapidly as functions of $N$
via our UV-based truncation than via the traditional IR-based truncation.
We can also see this from Fig.~\ref{fig:Nscaling}.
In Fig.~\ref{fig:Nscaling}, the blue curves illustrate 
the IR-based approach to asymptotia for these quantities.
However, in Fig.~\ref{fig:Nscaling} we have also
superimposed the red curves which represent the results of our  UV-based  
approach to asymptotia for these same quantities.
In all cases, we see that the UV-based approach to asymptotia tends to differ
significantly from the IR-based approach for small $N$.   
Indeed, as $N$ increases, we see that the finite-$N$ energy densities associated
with the IR-based approach tend to approach their asymptotic limits {\it from above}\/,
while the finite-$N$ energy densities associated with the UV-based approach tend to approach
these same asymptotic limits {\it from below}.\/ 
Nevertheless, as \il{N\to\infty},
we see that the UV-based truncation approaches 
asymptotia more rapidly (for smaller values of $N$) than does the IR-based truncation.
We also observe that the approach to asymptotia provided by the UV-based truncation
is monotonic, particularly for the late-time tower fraction $\etabar$,
whereas the path provided by the IR-based approach is non-monotonic.    However, this too is 
straightforward to understand.
In general, the non-monotonicity of 
$\etabar(N)$ in the IR-based approach is due to the somewhat spurious effects 
of the highest mass eigenstate in the finite-$N$ system, 
as discussed in Sect.~\ref{sec:TheModel} and sketched in 
Figs.~\ref{fig:massspectrum} and \ref{fig:initialabruptabundance}.  
In the UV-based approach, by contrast, the highest mode no longer behaves
anomalously for finite $N$.

\begin{figure}[b]
\includegraphics[keepaspectratio, width=0.5\textwidth]{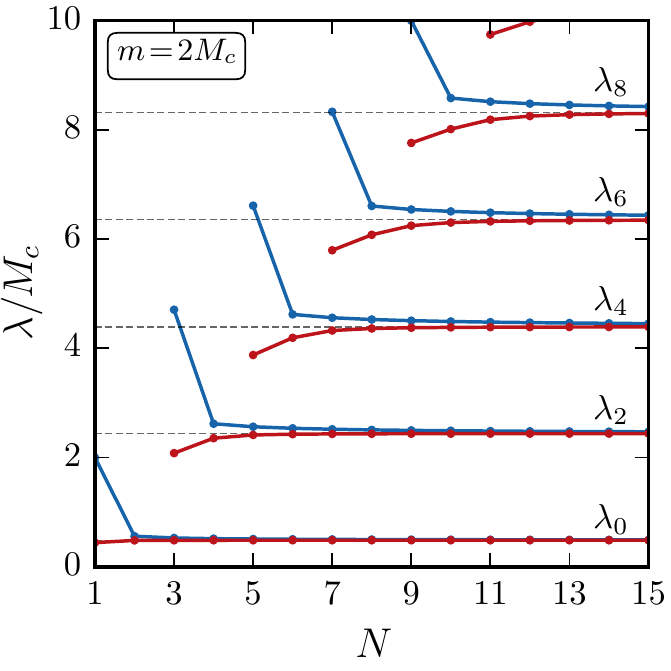}
\caption{Eigenvalues of the IR-based and UV-based mass matrices, plotted respectively in blue and red 
as functions of $N$ for \il{m/M_c=2}.  For visual clarity, only even eigenvalues are shown.
In each case 
the blue and red curves asymptote to the exact eigenvalues $\lambda_k$ as
\il{N\to\infty}, but the UV-based red curves generally approach this limit more rapidly and more smoothly 
than do the IR-based blue curves.
Note that each blue curve begins with an anomalously high eigenvalue;
this is nothing but the truncation artifact already sketched in Fig.~\protect\ref{fig:massspectrum}
for the highest mode $\lambda_{N-1}$ in each case. 
By contrast, such artifacts are entirely eliminated in the UV-based approach.  }
\label{fig:lambdascaling}
\end{figure}

These features are also readily apparent simply by comparing the eigenvalues 
of the IR-based truncated mass matrix $\calM^2_{k\ell}$ in Eq.~(\ref{kmassmatrix}) with those
of the UV-based truncated mass matrix $\widetilde \calM^2_{k\ell}$ in Eq.~(\ref{massmatrixnew}) as functions of $N$.
Our results are shown in Fig.~\ref{fig:lambdascaling} for \il{m/M_c=2}. Once again, we see that the 
asymptotic \il{N\to\infty} limit is approached more rapidly and more smoothly in the UV-based approach
(red curves) than in the IR-based approach (blue curves).   Moreover, we see that the anomalous highest 
eigenvalue which appears in the latter approach no longer exists in the former.

At the end of Sect.~\ref{road}, we introduced a multi-component adiabatic 
approximation~(\ref{eq:multiadiabaticapproximation}) and noted
that this quantity provides a lower bound on the total energy density of
our $N$-mode system because it includes only the contributions from the lightest mode.
Like other physical finite-$N$ quantities, however, the value of this quantity and its
interpretation as a lower limit depend 
on the specific KK truncation chosen, since 
the truncation in some sense determines what is meant by the lightest mode
and whether its contributions are affected by the removal of the 
higher modes.   
This is readily apparent in Fig.~\ref{fig:Nscaling}, where the red ``UV-based''
values of the total energy density are clearly 
smaller for certain small values of $N$ than the ``IR-based''
values of the adiabatic lower limit.
These red ``UV-based'' total energies nevertheless strictly exceed 
the values of a corresponding ``UV-based'' adiabatic lower limit. 

We conclude with a final comment.  Although we have presented our alternate UV-based 
truncation of the full KK theory as providing a more rapid road to asymptotia, the 
existence of such an alternate truncation is interesting in its own right. In any theory 
involving extra spacetime dimensions, one can never probe all energy scales and thereby 
detect all KK modes.  Instead, we expect the physics of our full KK system to be 
approximately represented at low energies through some sort of truncation that focuses 
on the lower modes. {\it However, if the physics of our extra dimensions results in a 
mixing of KK modes (as must always arise in any theory which breaks translational invariance 
in the extra compactified dimension), we now see that there are multiple options for 
performing such a truncation.}\/ Indeed, one could even argue that our UV-based truncation 
is more appropriate for certain calculations since it incorporates and thus is more sensitive 
to the actual masses of the physically propagating mass-eigenstates that we would expect 
to observe experimentally.  This last statement is of course subject to various 
renormalization-group effects which could potentially deform the observed KK masses and 
couplings, as discussed in Refs.~\cite{sky1,sky2,sky3}, and which exist even for theories such as those 
considered in Refs.~\cite{sky1,sky2,sky3} in which no KK-mixing is present. These observations nevertheless 
potentially give our UV-based truncation a theoretical importance in its own right that 
renders it worthy of further study.

Of course, there does exist a well-defined procedure through which one can unambiguously 
describe the physics of our full KK theory at low energies: one can use the methods of 
effective field theory (EFT).~ Specifically, one carefully integrates out the modes with energies 
above a particular cutoff scale, thereby obtaining not only a truncated KK tower exhibiting 
renormalized masses, but also a set of effective operators which reflect the underlying structure 
of the full theory.  In general, the structure of such an EFT can be quite complicated. 
Thus, it is traditional in the literature to simply adopt the IR-truncated theory as an 
approximation to this EFT.~ Our point, then, is that our UV-based truncation might profitably 
serve as an alternative approximation to the complete EFT --- an approximation which, as we have 
discussed, may have certain advantages.  Needless to say, the UV and IR truncations, as well as 
the complete EFT, all converge to the full KK theory as \il{N\rightarrow\infty}.

\FloatBarrier
\section{KK Tower Limit: ~\texorpdfstring{$N=\infty$}{N=infty}\label{sec:KKLimit}}


\begin{figure*}
\includegraphics[keepaspectratio, width=0.9\textwidth]{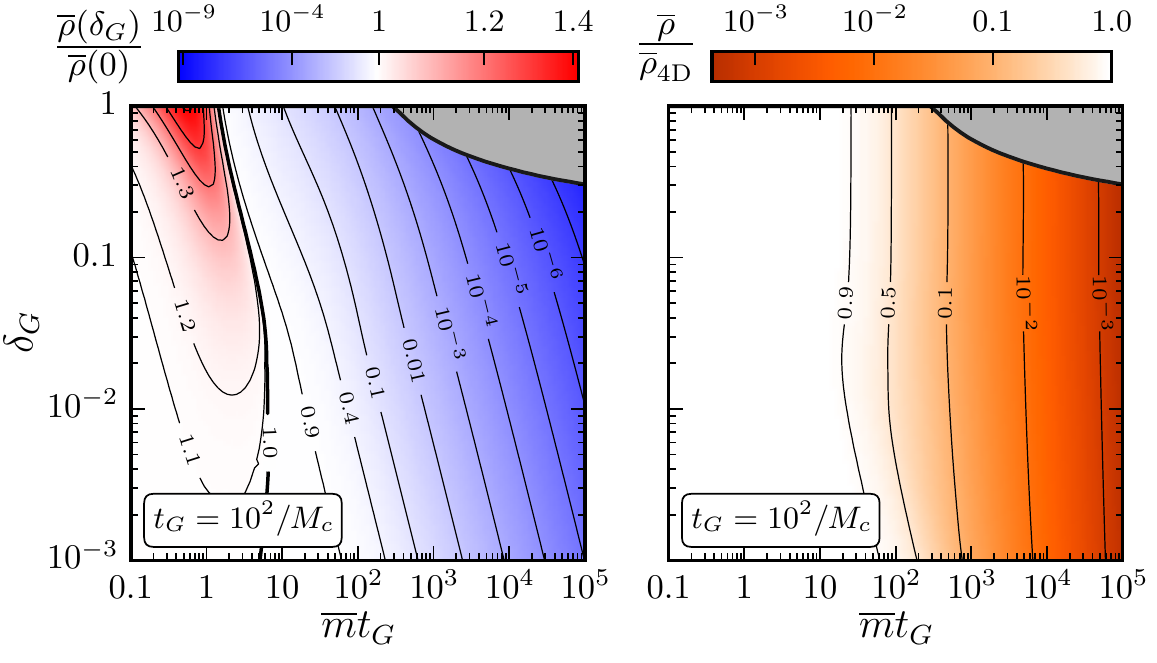}
\caption{The late-time energy density $\rhobar$ 
of the full KK tower, plotted in the \il{(\mbar t_G, \delta_G)} plane for \il{t_G= 10^2/M_c}.
The left and right panels show contours of 
$\rhobar/ \rhobar(\delta_G=0)$ and
$\rhobar/ \rhobar_\fd$, respectively.}
\label{fig:infinitemodedensityplots}
\end{figure*}

Having studied our system as a function of $N$ for large \il{N\gg 1}, 
we are now finally in a position to present our 
results for the late-time energy density $\rhobar$ and tower fraction $\etabar$
for the full, infinite KK tower. 
As we shall see, certain features emerge in the full infinite-$N$ limit
that were not present for any finite $N$.
Taken together, these results thus describe the late-time energy configuration
across our infinite KK tower when 
it has been subjected to a mass-generating phase transition 
of arbitrary width $\delta_G$ and arbitrary magnitude (set by the late-time brane mass $\mbar$)
at time $t_G$ during its cosmological history. 

We shall present our results in several different ways.
First, in Fig.~\ref{fig:infinitemodedensityplots},
we plot our results for the total late-time energy density $\rhobar$ as a fraction of
$\rhobar(\delta_G=0)$ (left panel)
or $\rhobar_\fd$ (right panel).
Our first observation 
from the results in Fig.~\ref{fig:infinitemodedensityplots}
is that while a quantity such as $\rhobar(\delta_G=0)$ 
might indeed be a useful approximation for $\rhobar$ which is 
valid in certain regions of parameter space, 
such an approximation can fail badly in others.
For example, we see that the abrupt (\il{\delta_G=0}) approximation works
well for small \il{\mbar t_G\lsim 0.1} even if $\delta_G$ is sizable but 
is otherwise capable of either significantly underestimating or overestimating the 
true late-time energy density $\rhobar$, the latter often by many orders
of magnitude.

Another immediate observation is that the contours in
Fig.~\ref{fig:infinitemodedensityplots}
follow a power-law type behavior in those
regions of \il{(\mbar t_G,\delta_G)} parameter space in which we are eventually going to be the
most interested,
namely those regions with
\il{\mbar t_G\gsim 10} and \il{\delta_G\lsim 0.3}
in which multiple components contribute non-trivially to the total late-time abundance and \il{\etabar>0}. 
This power-law behavior enables us to extract approximate analytical expressions for
$\rhobar/ \rhobar(\delta_G=0)$ and
$\rhobar/ \rhobar_\fd$ which are valid in these regions.
Specifically, given the results in Fig.~\ref{fig:infinitemodedensityplots}
(as well as analogous results calculated for different values of $t_G$),
we find
\beq\label{eq:simplewidthnormapproximation}
\frac{\rhobar(\delta_G)}{\rhobar(0)} ~\approx~ \frac{0.6}{\delta_G}\, \frac{e^{-\frac{1}{2} M_c/\mbar}}{\mbar t_G}
         \left(\frac{M_c}{\mbar}\right)^{0.9}
\eeq
and
\beq\label{eq:mixingnormformula}
  \frac{\rhobar}{\rhobar_{\rm 4D}} ~\approx~ \frac{1}{1 + 2(\mbar/M_c) e^{-\frac{1}{2}M_c/\mbar }} \ .
\eeq
While the expression in Eq.~(\ref{eq:simplewidthnormapproximation})  holds 
to within $\pm 25\%$ across the relevant \il{\etabar>0} region,
this result is actually somewhat sensitive to the value of $t_G$, with the upper error limit 
growing smaller with increasing $t_G$ and bigger with decreasing $t_G$.
By contrast, the expression in Eq.~(\ref{eq:mixingnormformula}) holds to within $\pm 5\%$ across the relevant
region, making it one of the most accurate analytical approximations we have presented in this paper.

We can also deduce approximate scaling laws from these expressions.
For \il{\mbar/M_c \gg 1}, we find that
Eq.~(\ref{eq:simplewidthnormapproximation}) approximately reduces to
\beq
    \frac{\rhobar(\delta_G)}{\rhobar(0)} ~\sim~ {1\over \mbar t_G \delta_G}  \left( {M_c\over \mbar}\right)~
\label{scaling1}
\eeq
which is a factor of $M_c/\mbar$
greater than the corresponding expression we observed in Eq.~(\ref{fdscaling})
for the 4D case with \il{N=1}.
Thus the presence of an entire of tower of KK states
suppresses what would otherwise have been
the late-time energy density of the zero mode alone
by an additional factor of $M_c/\mbar$.
Likewise, for \il{\mbar/M_c\gg 1},
we observe from
Eq.~(\ref{eq:mixingnormformula})
that
\beq
    \frac{\rhobar}{\rhobar_\fd} ~\sim~ {M_c\over \mbar}~.
\label{scaling2}
\eeq
Thus, combining these results for \il{\mbar/M_c\gg 1}, we find that
\beq
         {\rhobar(\delta_G) \over \rhobar(0)} ~\sim~
         {\rhobar(\delta_G) \over \rhobar_\fd(\delta_G)} \cdot
         {\rhobar_\fd(\delta_G) \over \rhobar_\fd(0)}~
\eeq
from which we deduce that 
\beq
      \rhobar(0) ~\sim~ \rhobar_\fd(0)~.
\label{same}
\eeq
This in turn requires that there exist a constant $c$ for which
\beq
           \lim_{\delta_G\to 0} \left( {\rhobar\over \rhobar_\fd} \right) ~\approx ~c~,
\label{limitratio}
\eeq
and indeed this last relation is true for \il{c=1}. 
According to the results shown in the right panel of Fig.~\ref{fig:infinitemodedensityplots},
this relation with \il{c=1} is manifestly true for \il{\mbar/M_c \lsim 1}.  
However, as $\delta_G$ approaches zero, this relation with \il{c=1} becomes true
for larger and larger values of $\mbar/M_c$.
Thus, all of the scaling relations we have quoted here are self-consistent
within the regions of validity claimed.

In this connection, we remark that it is not only the power-law scaling relations
in Eqs.~(\ref{scaling1}) and (\ref{scaling2}) 
which must be consistent with each other;
the same must also be true of the more complete expressions such as those
in Eqs.~(\ref{eq:simplewidthnormapproximation})
and (\ref{eq:mixingnormformula}) from which these power-law relations are derived.
However, although such self-consistent pairs of expressions exist,
the specific expressions provided in
Eqs.~(\ref{eq:simplewidthnormapproximation})
and (\ref{eq:mixingnormformula}) 
do not constitute such a pair.  Rather, these expressions are provided instead because they 
yield even greater numerical accuracy over their appropriate regions of
parameter space.

\begin{figure*}[t]
\includegraphics[keepaspectratio, width=1.0\textwidth]{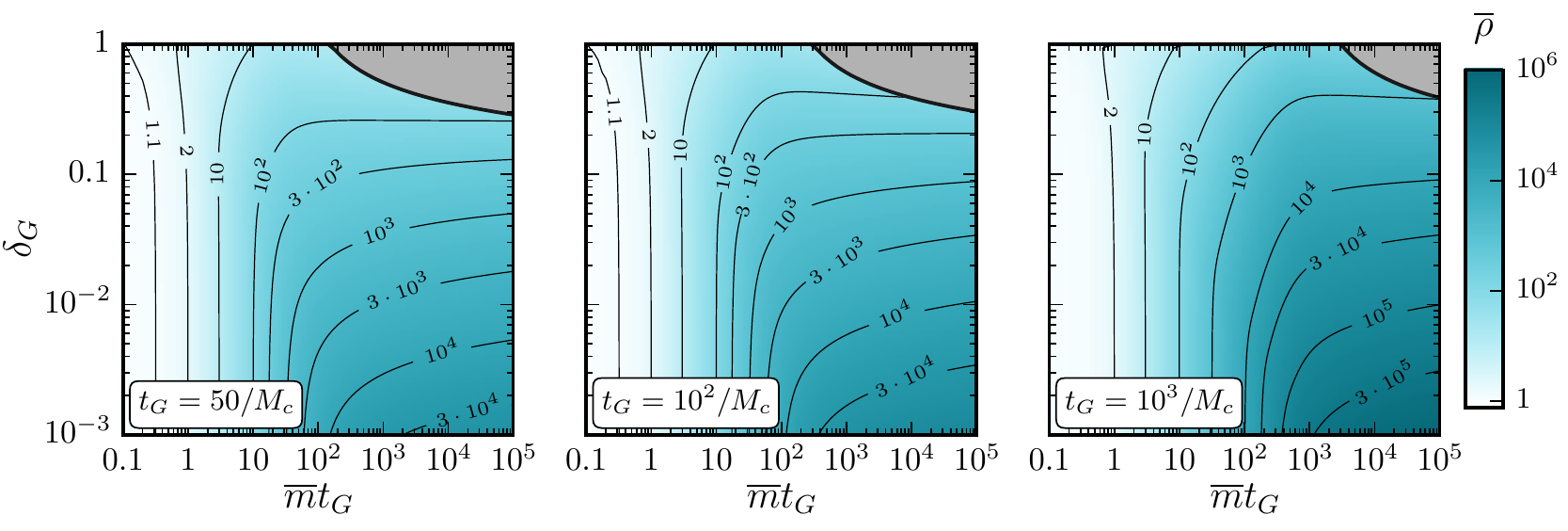}
\caption{The {\it absolute}\/ late-time energy density $\rhobar$ 
in units of $\half \langle \phi\rangle^2 t^{-2}$, plotted within the $(\mbar t_G,\delta_G)$ 
plane for three different values of $M_c t_G$.}
\label{fig:absoluteplots}
\end{figure*}

As we have seen, considering the late-time energy
density $\rhobar$ as a fraction of related benchmarks such as $\rhobar(\delta_G=0)$ or $\rhobar_\fd$ 
has been useful for understanding the {\it relative}\/ effects of increasing $\delta_G$ or 
increasing the numbers of modes.
However, we also wish to understand the late-time energy densities $\rhobar$ on an {\it absolute}\/ scale. 
Of course, 
these absolute magnitudes can in principle be obtained through a sequence of multiplications of previous 
intermediate results.
For example, one possible path is given by
\beq
           \rhobar ~=~ \left({\rhobar \over \rhobar_\fd}\right) 
        \, 
            \left({\rhobar_\fd \over \rhobar_\fd(\delta_G=0)}\right) 
        \,  
            \rhobar_\fd (\delta_G=0)~
\eeq
where these three different factors are indicated in the left panel of Fig.~\ref{fig:infinitemodedensityplots},
the left panel of Fig.~\ref{fig:singlefielddensityplot}, and 
the large-$t$ behavior of Eq.~(\ref{eq:abruptapproximation}) respectively.
However, since each of these quantities is generally a complicated function of $\mbar t_G$ and $\delta_G$ 
throughout the $(\mbar t_G,\delta_G)$ parameter space,
it is not readily apparent what composite behavior might emerge from these individual factors.
It is therefore useful to compile our intermediate results together in order to present the resulting 
values for $\rhobar$ as full, absolute quantities.

Our results are shown in Fig.~\ref{fig:absoluteplots} for three different values of $t_G$.
In general, we see that taking larger values of $\mbar t_G$ results in larger values of $\rhobar$.
However, we also see that taking larger values of $\delta_G$ tends to suppress $\rhobar$.
Together, these effects conspire to produce the curved contours shown.
Indeed, increasing $t_G$ relative to $M_c$ then tends to push these contours to the right, thereby
again increasing the late-time energy density $\rhobar$ still further.

\begin{figure}[t]
\includegraphics[keepaspectratio, width=0.49\textwidth]{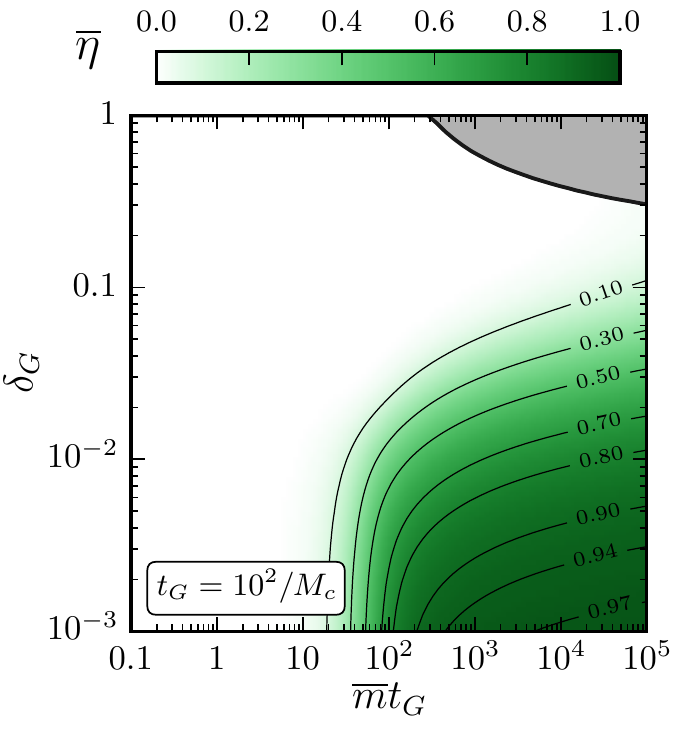}
\caption{The late-time tower fraction $\etabar$ of the KK tower,
plotted within the \il{(\mbar t_G,\delta_G)} plane for \il{t_G=10^2/M_c}.
Unlike the tower fractions illustrated in Fig.~\ref{fig:finitemodedensityplotsgenN2} for finite $N$,
we see that the tower fraction for \il{N=\infty} 
is now monotonic in both $\mbar t_G$ and $\delta_G$
and approaches unity for large $\mbar t_G$ and small $\delta_G$.}
\label{fig:infinitemodeeta}
\end{figure}

In Fig.~\ref{fig:absoluteplots} we indicated the 
full, absolute magnitudes of $\rhobar$ for entire our KK tower as functions of $\mbar t_G$ and $\delta_G$.
However, we are also interested in 
the distribution of this total energy density across the different modes of our KK tower.
As we have seen throughout this paper,
one measure of this distribution is the tower fraction $\eta$ --- the 
fraction of the total abundance which is carried by all but the most abundant mode in the tower.
The late-time values of these tower fractions for the full KK tower are plotted 
in Fig.~\ref{fig:infinitemodeeta}.
As we see upon comparison with its finite-$N$ 
equivalents in Fig.~\ref{fig:finitemodedensityplotsgenN},
 {\it passing to the infinite-$N$ limit has the effect of removing all non-monotonicities in $\etabar$}.
Indeed, we see that $\etabar$ now increases monotonically as either $\mbar t_G$ is increased or $\delta_G$ is 
decreased.   We stress that this is a feature which emerges only for the full KK tower with \il{N=\infty},
but which would not be true for any finite value  of $N$.   
As a result, {\it it is the region with large $\mbar t_G$ (or equivalently large $\mbar/M_c$) and relatively
small $\delta_G$ for which the total energy density ends up distributed most broadly across the different
states in the KK tower at late times.}

\begin{figure*}[t]
\includegraphics[keepaspectratio, width=0.8\textwidth]{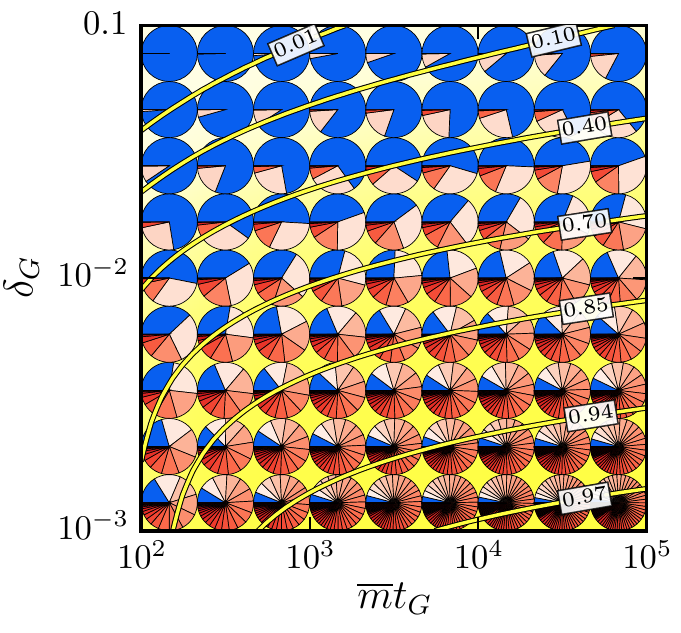}
\caption{A graphical representation of the distribution of the
energy densities $\rhobar_\lambda$ across the entire KK tower, calculated for a
square ``lattice'' of locations
in the \il{(\mbar t_G,\delta_G)} plane.
Each pie chart 
indicates how the total $\rhobar$ is distributed across the different KK modes,
with the blue slices indicating the contribution associated with the lightest mode
and the remaining slices (colored pink through dark red) indicating
the contributions from successively heavier KK modes.
The results for each pie correspond to the parameters \il{(\mbar t_G,\delta_G)}
associated with the location of the center of the pie,
and the yellow contours indicate the corresponding values of $\etabar$,
as taken from Fig.~\protect\ref{fig:infinitemodeeta}.
We see that the total energy density is preferentially captured by the lightest
mode for larger $\delta_G$ or smaller $\mbar t_G$, but that the total energy density
is distributed more democratically across the KK tower for smaller $\delta_G$ and larger $\mbar t_G$.
Thus a wide variety of energy-density distributions across the KK tower can be
realized simply by adjusting the parameters associated with the mass-generating phase transition.}
\label{fig:rhofracpies}
\end{figure*}

While $\etabar$ represents one measure of the degree to which the late-time total energy density of the KK tower
is distributed across its different modes, 
this quantity still does not tell us how many modes are actually
carrying a significant abundance.
More specifically, we would like to know   the shape the overall ``profile'' of the 
energy-density distribution across the entire KK tower as a function of $\mbar t_G$ and $\delta_G$.
Towards this end, can visualize the shape of a given energy-density distribution profile by means of 
a pie chart whose different pie slices illustrate how $\rhobar$ is distributed across the different KK modes.
We then seek to understand how the relative slices of such pies 
evolve as functions of $\mbar t_G$ and $\delta_G$.

Our results are shown in Fig.~\ref{fig:rhofracpies}  for \il{t_G=10^2/M_c}.
In Fig.~\ref{fig:rhofracpies}
we have focused on the non-trivial green region of Fig.~\ref{fig:infinitemodeeta} wherein \il{\etabar>0},
and then superimposed a set of pie charts illustrating how the energy-density profile varies
across this region. 
We see from this figure that the total energy density tends to be preferentially captured by the lightest
mode as $\delta_G$ increases or $\mbar t_G$ decreases, but that 
the total energy density tends to be more democratically distributed across the KK modes of the tower
otherwise.  {\it We see, then, that a wide variety of energy-density distributions across the KK tower are possible
and can be realized simply by adjusting the parameters associated with the mass-generating phase transition.}

\begin{figure*}[t]
\includegraphics[keepaspectratio, width=0.8\textwidth]{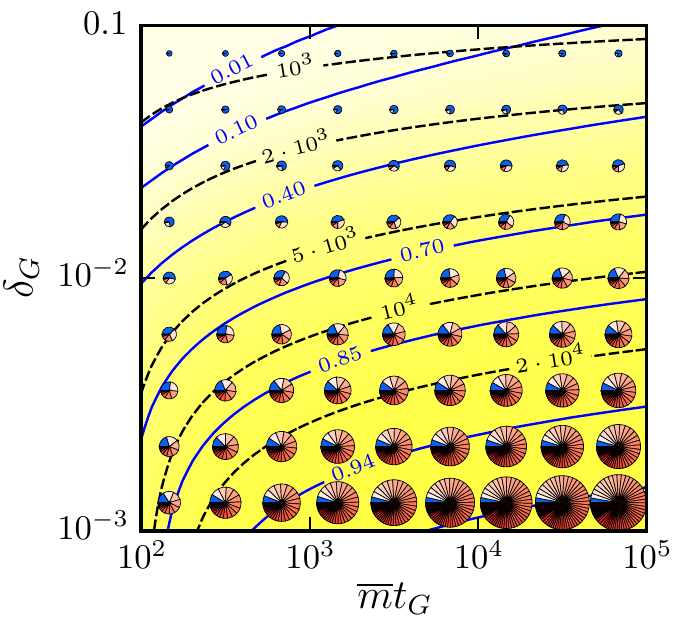}
\caption{A grand summary of the results of this section, combining 
our information concerning the absolute {\it magnitudes}\/ of the total late-time
energy density $\rhobar$ from Fig.~\protect\ref{fig:absoluteplots} 
with our information concerning the {\it distribution}\/ of that energy density from
Fig.~\protect\ref{fig:rhofracpies}, all plotted 
within the \il{(\mbar t_G, \delta_G)} plane.
This figure is essentially the same as Fig.~\protect\ref{fig:rhofracpies} except that the overall
areas of our pie charts have been rescaled in proportion to the magnitudes of the total late-time
energy densities $\rhobar$ shown in Fig.~\protect\ref{fig:absoluteplots}.
In the background we have also indicated the contours of $\rhobar$ from Fig.~\protect\ref{fig:absoluteplots}
(dashed black lines)
as well as the contours of $\etabar$ from Fig.~\protect\ref{fig:infinitemodeeta} (solid blue lines).
We see that there is a non-trivial {\it correlation}\/ between the overall  magnitude of $\rhobar$
and the distribution of that energy density, with larger energy densities distributed more
democratically across the KK tower and smaller energy densities captured more and more preferentially
by the lightest KK mode.   Both features are nevertheless extremely sensitive to the parameters governing
the mass-generating phase transition, with the former behavior dominating for smaller $\delta_G$ and larger
$\mbar t_G$ and the latter 
dominating for larger $\delta_G$ and smaller $\mbar t_G$.} 
\label{fig:rhofracpiesgrand}
\end{figure*}

Finally, we can summarize the results of this section by combining our 
information concerning the absolute {\it magnitudes}\/ of the late-time total energy density $\rhobar$,
as indicated in Fig.~\ref{fig:absoluteplots},
with our information concerning the late-time {\it distributions}\/ of that total energy density,
as indicated in Fig.~\ref{fig:rhofracpies}.
To do this, we can begin with the information in Fig.~\ref{fig:rhofracpies} but then 
{\it rescale the size of each pie chart so that 
the area of each pie chart is proportional to the total energy density $\rhobar$.}
The result is shown in Fig.~\ref{fig:rhofracpiesgrand}.

One important lesson that emerges from 
Fig.~\ref{fig:rhofracpiesgrand} is that
there is a non-trivial correlation between the overall {\it magnitude}\/ of $\rhobar$
and the {\it distribution}\/ of that energy density, with larger energy densities distributed more
democratically across the KK tower and smaller energy densities captured more and more preferentially
by the lightest KK mode.   
The results in Fig.~\ref{fig:rhofracpiesgrand} allow us to see rather dramatically
the effects of our mass-generating phase transition on the eventual late-time energy configuration
of our KK tower.
For any fixed $\mbar t_G$, we see that increasing the phase-transition timescale $\delta_G$ 
suppresses the energy density that remains in the massive KK modes, causing the lighter
modes to assume an increasing fractional share of the total energy density.
Yet, at the same time, increasing $\delta_G$ suppresses the total energy
density that is ultimately pumped into the KK tower by the phase transition.
Likewise, for any $\delta_G$, we see that increasing $\mbar t_G$ 
increases the total energy pumped into the system by the phase transition
while simultaneously causing this energy density to be more democratically distributed.
{\it Thus, merely by choosing appropriate values of $\mbar t_G$ and $\delta_G$, it is possible to adjust the 
total absolute energy density remaining in the KK tower at late times to any value one might
select for phenomenological purposes while simultaneously retaining the ability to adjust
the distribution of that energy density across the different KK modes.}   This, then, is the power
of the mass-generating phase transition and the influence of its associated timescale.

\FloatBarrier
\section{Example:~ Axion in the Bulk\label{sec:AxionintheBulk}}


Until this point, we have maintained generality
by considering a higher-dimensional field $\Phi$
and specifying little more about this field than that it is a scalar.
Likewise, we have assumed little more about our phase transition
than that it generates masses in a time-dependent way.
However, in order to explore one possible set of phenomenological implications of our results,
we shall now apply our machinery to the case in which $\Phi$ is an axion-like
particle and in which our phase transition on the brane is one in which 
instanton-like effects give mass to that axion.

We begin by considering the setup
described in Refs.~\cite{DDM2,DDMAxion}, which is itself a generalization of
an earlier framework considered in Ref.~\cite{DDGAxions}.  
Specifically, 
we consider the same five-dimensional geometry as discussed in previous sections
and henceforth take 
\beq
    M_c ~=~ 4.49 \,\times\, 10^{-12} ~\text{GeV} 
\eeq
as our compactification scale.
This value of $M_c$ corresponds to \il{R\approx 44~\mu\text{m}}, 
which is the largest flat extra dimension 
allowed according to data from torsion-balance experiments~\cite{KapnerEotvosExpt}. 
Within the bulk of this extra dimension 
we shall consider
a (pseudo-)scalar field $\Phi$ which is
a straightforward generalization of the traditional QCD axion~\cite{PecceiQuinn1,WeinbergAxion,WilczekAxion}.
In particular,
we shall take $\Phi$ to be 
the Nambu-Goldstone boson associated with a global chiral Peccei-Quinn-like $U(1)_X$ symmetry which is spontaneously 
broken at some scale $f_X$.  
This $U(1)_X$ symmetry is assumed anomalous, and for a non-Abelian gauge group $G$ on the brane
with coupling $g$ and field strength $\mathcal{G}_{\mu\nu}$, this anomaly therefore generates a topological 
brane term of the form
\beq
\mathcal{L}_{\text{brane}} ~\rightarrow~  
\mathcal{L}_{\text{brane}} +
\frac{\mathcal{C} g^2}{32\pi^2}\frac{\Phi}{f_X^{3/2}}\text{Tr }\mathcal{G}_{\mu\nu}\tilde{\mathcal{G}}^{\mu\nu} 
\eeq
where $\mathcal{C}$ is a model-dependent constant. 
While this term has no effect on the classical 
equations of motion, it affects the vacuum structure of the theory. 
As the universe cools 
and reaches \il{T\sim \Lambda_G},
where $\Lambda_G$ is the confinement scale associated with the group $G$, 
instanton effects on the brane
explicitly break the $U(1)_X$ axion shift symmetry that prevents the axion from acquiring a mass.
As a result, a small temperature-dependent axion mass $m_X(T)$ is generated on the brane. 
In general, we shall model the time dependence of this mass 
exactly as in previous sections, as resulting from a phase transition 
occurring at a time $t_G$ with a width $\delta_G$.
In this connection we note that our adoption of an LTR cosmology,
as discussed in Sect.~\ref{sec:TheModel}, 
implies that this phase transition takes place 
during an inflaton-dominated (and thus matter-dominated) epoch.
The corresponding  time/temperature relationship then yields
the result
\beq
t_G ~=~ \sqrt{\frac{45 g_*(T_{\rm RH})}{2\pi^2}}\frac{T_{\rm RH}^2 M_p}{g_*(\Lambda_G)\Lambda_G^4} ~
\label{timetemp}
\eeq
where \il{T_{\rm RH}\sim\mathcal{O}(\text{MeV}) < \Lambda_G} is the reheating temperature, 
where $g_*(T)$ is the effective number of relativistic degrees of freedom at temperature $T$, 
and where \il{M_p\equiv 1/\sqrt{8\pi G}} is the reduced Planck mass. 
For concreteness we shall take \il{T_{\rm RH}= 30}~MeV in what follows.
Finally, at late times, our phase transition leaves our axion-like field with
a brane mass given by
\beq
\mbar_X^2 ~\equiv~ \frac{\mathcal{C}^2 g^2}{32\pi^2}\frac{\Lambda_G^4}{\fhX^2}~ 
\label{mbarXdefn}
\eeq
where we have defined the effective four-dimensional $U(1)_X$-breaking 
scale \il{\fhX\equiv \sqrt{2\pi R f_X^3}}. 

The above setup defines our five-dimensional axion theory.
Compactifying this theory via KK reduction  
then yields an effective four-dimensional theory consisting of a KK tower of axion modes $\phi_k$ 
whose mass matrix takes exactly the form given in Eq.~(\ref{kmassmatrix}), with $m(t)$ now 
identified as $m_X(t)$. 
Thus, we see that our axion model is nothing but a special case of the model we have already considered
thus far --- the special case in which we identify $\mbar$ as $\mbar_X$ and identify $t_G$ as the cosmological time 
corresponding to the temperature \il{T\sim\Lambda_G}, as in Eq.~(\ref{timetemp}).

In previous sections, we studied the behavior of our general KK system 
as a function of the four parameters \il{\lbrace \mbar, M_c, t_G, \delta_G\rbrace}.
For our axion theory, by contrast, we see that 
\il{\mbar\to \mbar_X}, and moreover we see that both $\mbar_X$ and $t_G$ are themselves
related to the more fundamental parameters $\Lambda_G$ and $\hat f_X$ through 
Eqs.~(\ref{mbarXdefn}) and (\ref{timetemp})
respectively.
Thus, we shall henceforth consider our axion theory 
to be parametrized by
\il{\lbrace \Lambda_G, \hat f_X, M_c, \delta_G\rbrace}.

In four dimensions, the methods used to estimate the late-time abundance of a given axion field typically fall into one
of two classes examined in Sect.~\ref{sec:FourDimensionalLimit}: the adiabatic approximation
or the abrupt approximation.
As we have seen, the former only applies when the mass is generated sufficiently slowly during field oscillations,
with \il{\dot m \ll m^2}, 
while the latter applies only for very small $\delta_G$ (such that the phase-transition width $\Delta_G$ is much
smaller than the timescale of field oscillations) or for situations in which \il{t_\zeta\gg t_G} (so that the phase
transition occurs while the field is still overdamped and essentially frozen).
However, as we have seen in previous sections, these approximations break down in
relatively large regions of parameter space.
We should also expect deviations from our standard expectations in the case of
an infinite KK tower of axion modes.
For example, the virialization condition that underpins the adiabatic approximation
now becomes 
 \il{\dot{\lambda}_0 \ll \lambda_0^2}, and $\lambda_0$ can be much smaller than
$m_X$~\cite{DDGAxions}.
Likewise, there is always some subset of modes   
in the KK tower for which \il{t_\zeta^{(\lambda)}\lesssim t_G}.
Such modes are therefore necessarily affected by the time-dependence of the phase transition.

\begin{figure*}
\includegraphics[keepaspectratio, width=1.0\textwidth]{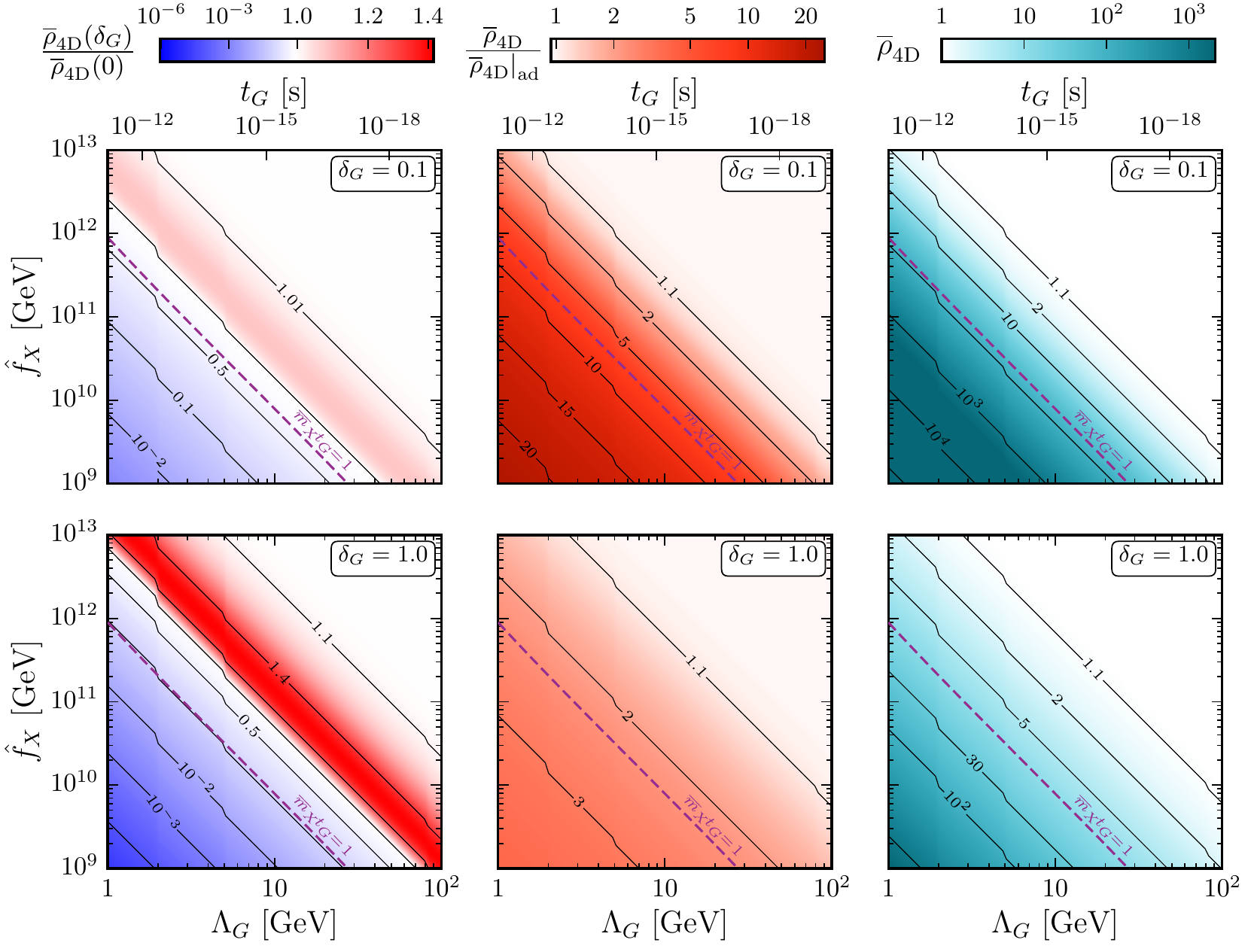}
\caption{ The late-time energy density $\rhobar_\fd$ of our generalized axion field in the 4D limit, 
plotted within the \il{(\Lambda_G, \hat f_X)} plane for \il{\delta_G = 0.1} (top row) and
\il{\delta_G=1.0} (bottom row).   Note that the horizontal $\Lambda_G$ axis (as shown along the bottom of each panel)
is equivalently a $t_G$ axis (as shown along the top).
The panels in the left and middle columns plot
$\rhobar$ as fractions of the abrupt and adiabatic approximations $\rhobar_\fd(\delta_G=0)$ 
and $\rhobar_\fd|_{\rm abs}$, respectively,
while the panels in the right column 
plot the absolute magnitude of $\rhobar_\fd$ in units of    
$\half \langle \phi\rangle^2 t^{-2}$.   
Also shown in each panel is a purple dashed line indicating the contour
along which \il{\mbar_X t_G=1}.}   
\label{fig:4Daxion}
\end{figure*}

In this section,
we shall therefore present exact results for the late-time energy densities of our axion fields.  We shall do this both
for the four-dimensional \il{N=1} case as well as the infinite-$N$ case of our full KK axion tower.
As discussed above, we shall take our axion parameter
space to be parametrized by
\il{\lbrace \Lambda_G, \hat f_X, M_c, \delta_G\rbrace}.
However, we stress that the plots to be presented in this section are not merely translations of our previous
plots into these new variables.   First, we have extended our range of 
interest within this parameter space into those regions of specific interest for axion physics.
Second, we shall now regard $\Lambda_G$ and $\hat f_X$ as the fundamental axion parameters
relative to which we wish to consider continuous variations. In other words,
we shall plot our the late-time energy densities $\rhobar$ 
as functions of $\Lambda_G$ and $\hat f_X$ within the \il{(\Lambda_G,\hat f_X)} plane, choosing only 
discrete representative choices for $\delta_G$.    This too is different than what was done in previous sections,
and thus represents a different, independent slice through our four-dimensional parameter space.  

\begin{figure*}[t]
\includegraphics[keepaspectratio, width=0.7\textwidth]{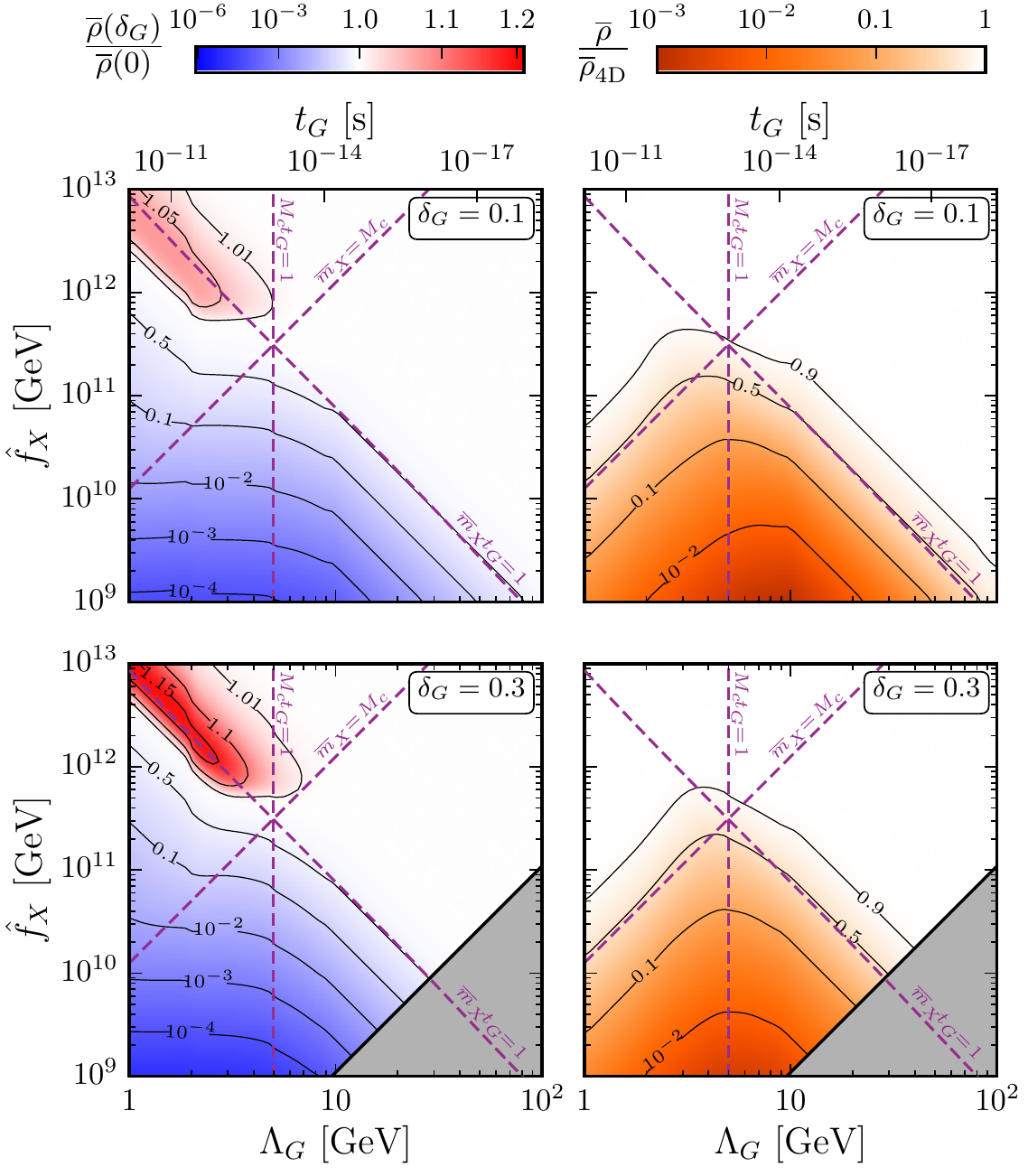}
\caption{ The total late-time energy density $\rhobar$ of our generalized axion KK tower, 
plotted as fractions of $\rhobar(\delta_G=0)$ (left column) or $\rhobar_\fd$ (right column)
within the \il{(\Lambda_G, \hat f_X)} plane for \il{\delta_G = 0.1} (top row) and \il{\delta_G=0.3} (bottom row).   
As in Fig.~\protect\ref{fig:4Daxion},
the $\Lambda_G$ axis is shown along the bottom of each panel
while the equivalent $t_G$ axis is shown along the top.
Also shown in each panel are three purple dashed lines indicating the contours along which 
\il{\mbar_X t_G=1}, \il{\mbar_X = M_c}, or \il{M_c t_G=1}.
As with other figures in this paper, the gray regions 
are excluded for the reasons discussed in the paragraph below Eq.~(\ref{bigwidth}).}
\label{fig:KKaxion1}
\end{figure*}

We begin, as in Sect.~\ref{sec:FourDimensionalLimit}, by considering the four-dimensional \il{N=1} limit.
In this case, we find the results shown in Fig.~\ref{fig:4Daxion}.
We observe from the panels in the left and center columns of 
Fig.~\ref{fig:4Daxion} that there are distinct regions
of parameter space in which the abrupt and adiabatic approximations fail
to model the true late-time energy density, with the adiabatic approximation
consistently underestimating the true energy density and the abrupt approximation
either under- or overestimating the true energy density, in some cases by many orders
of magnitude.   We also see that increasing either the confinement scale $\Lambda_G$   
or the $U(1)_X$ symmetry-breaking scale $\hat f_X$  
generally decreases the late-time energy density of our axion field.  
Increasing the width $\delta_G$ of our axion-induced phase transition also has the same effect.
Indeed, we see from the relatively straight contours in Fig.~\ref{fig:4Daxion} that
the absolute magnitude of the late-time energy density scales approximately as
\beq 
           \rhobar_\fd ~\sim~ 1/(\Lambda_G^4 \hat f_X^2)~,
\eeq
where the constant of proportionality is a non-trivial function of $\delta_G$.
This scaling behavior is of course consistent with the analogous result in Eq.~(\ref{4Dscaling}).  
Finally, we note that all of the contour lines in Fig.~\ref{fig:4Daxion}  
experience slight ``ripples'' at \il{\Lambda_G\approx 2}~GeV and $5$~GeV.
These ripples are physical, and correspond to the energy scales $\Lambda_G$ at which there
are changes in the number $g_\ast(\Lambda_G)$  of relativistic degrees of freedom (the former
corresponding to the threshold for the charm quark and tau lepton, and the latter corresponding to the threshold for the bottom quark). 

\begin{figure*}[t]
\includegraphics[keepaspectratio, width=1.0\textwidth]{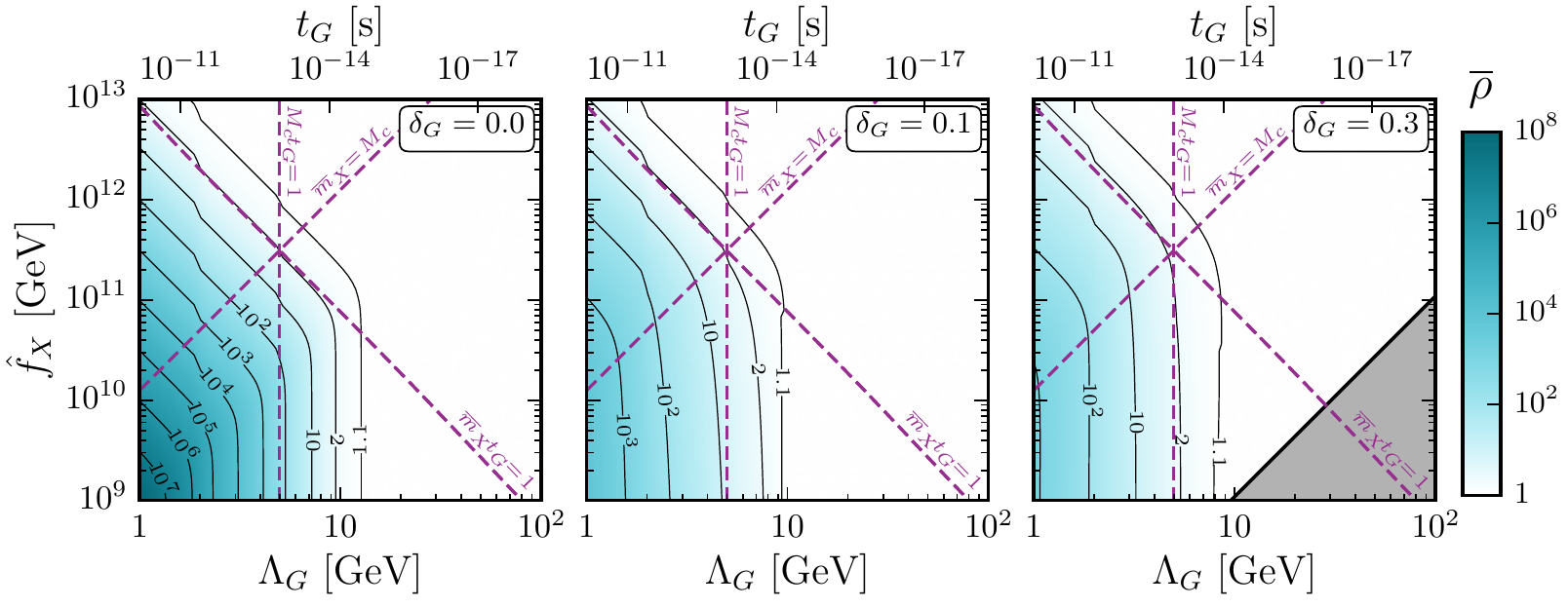}
\caption{ The absolute magnitude of the total late-time energy density $\rhobar$ of our generalized axion KK tower, 
plotted in units of $\half \langle \phi\rangle^2 t^{-2}$
within the \il{(\Lambda_G, \hat f_X)} plane for 
\il{\delta_G=0} (left panel),
\il{\delta_G = 0.1} (middle panel),
and \il{\delta_G=0.3} (right panel).  }   
\label{fig:KKaxion2}
%
\bigskip
\bigskip
\includegraphics[keepaspectratio, width=1.0\textwidth]{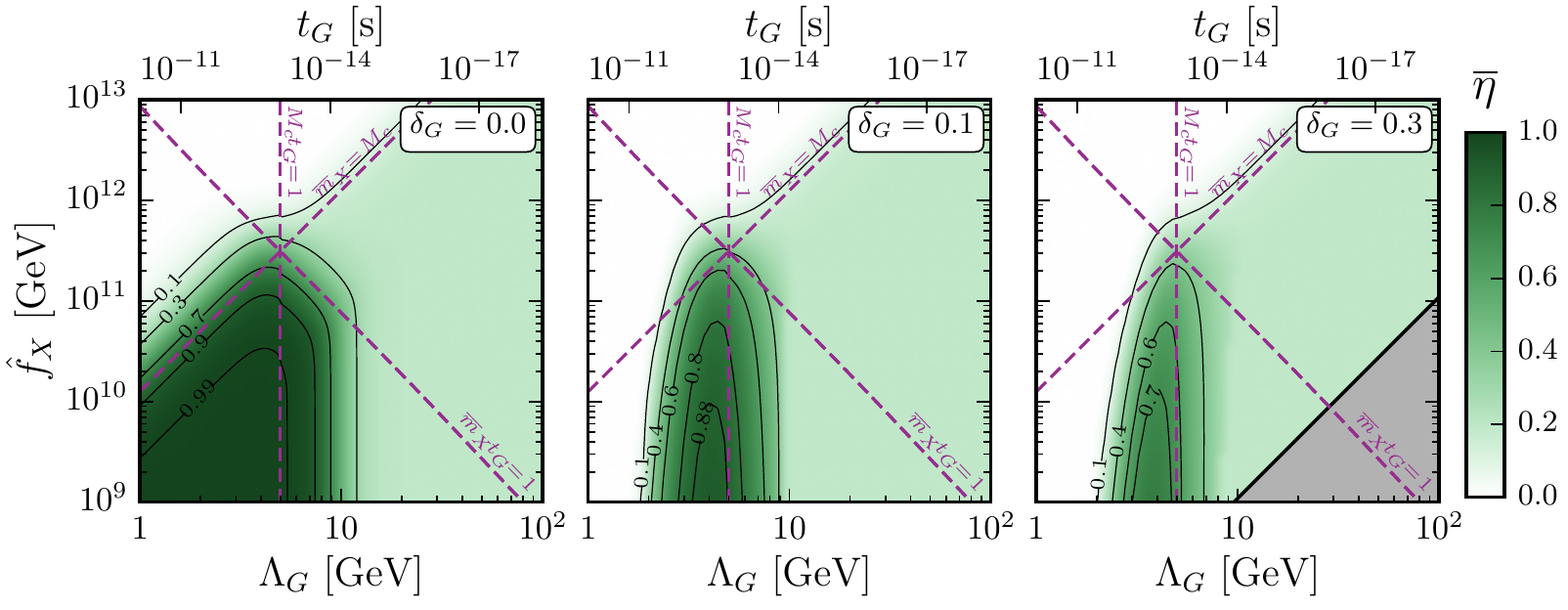}
\caption{ The late-time tower fraction $\etabar$ of our generalized axion KK tower,
plotted within the \il{(\Lambda_G, \hat f_X)} plane for 
\il{\delta_G=0} (left panel),
\il{\delta_G = 0.1} (middle panel),
and \il{\delta_G=0.3} (right panel).  }   
\label{fig:KKaxion3}
\end{figure*}

We now turn to the case in which a full KK tower of axion modes experiences
the instanton-induced phase transition.
In this case, our results are plotted in Figs.~\ref{fig:KKaxion1} through \ref{fig:KKaxion3}. 
In Fig.~\ref{fig:KKaxion1} 
we have plotted the values of the late-time total energy density $\rhobar$ of the KK tower
relative to our usual benchmarks $\rhobar(\delta_G=0)$ and $\rhobar_\fd$.
Once again, we see that the introduction of a non-zero width $\delta_G$ for our
instanton-induced phase transition leads to either enhancements or suppressions
in $\rhobar$, depending on the specific region of parameter space.
Moreover, we see that increasing $\delta_G$ only makes these enhancements or suppressions
more severe.   By contrast, turning to $\rhobar/\rhobar_\fd$, we see the introduction of the extra KK modes in the tower
only suppresses $\rhobar$ by an amount that grows {\it less}\/ severe for increasing $\delta_G$.
In all cases, however, these suppressions tend to be more pronounced for 
smaller $\Lambda_G$ and smaller $\hat f_X$ than they are in other regions of parameter space. 
It is worth noting that for smaller $\Lambda_G$ (but \il{\mbar_X \gg M_c}), we see a scaling behavior
of the form
\beq
     {\rhobar\over \rhobar_\fd} ~\sim~ {\hat f_X\over \Lambda_G^2}~. 
\eeq
However, as $\Lambda_G$ increases (and thus $t_G$ decreases),
this scaling behavior eventually breaks down 
and the behavior of $\rhobar/\rhobar_\fd$  
is increasingly determined solely by the behavior of the denominator $\rhobar_\fd$, with
contours of constant $\mbar_X t_G$.
Otherwise,
outside the $\mbar_X \gtrsim M_c$ and \il{\mbar_X t_G \gtrsim 1} region,
the energy density at late times is very similar to that in the 4D limit.

It is also instructive to compare our KK results for \hbox{$\rhobar/\rhobar(\delta_G=0)$} along the left column of
Fig.~\ref{fig:KKaxion1} with the analogous 4D results along the left column of Fig.~\ref{fig:4Daxion}.
For the full KK tower, we no longer find contours of constant $(\mbar_X t_G)^2$;  instead we find a region 
exhibiting contours with approximately constant $\hat f_X$.
Likewise, the enhancement region that we found in the 4D limit spanned the entire
\il{\lbrace \Lambda_G, \hat f_X\rbrace} parameter space for \il{\mbar_X t_G \sim 1}, whereas this region 
for the KK tower exists only for \il{\mbar_X\lsim M_c}. 

Next, we turn to Fig.~\ref{fig:KKaxion2} where we plot the absolute magnitude of the 
total late-time energy density of the KK axion tower in units
of $\half \langle \phi\rangle^2 t^{-2}$.
To help understand the features of these plots and how they evolve as a function of $\delta_G$, we 
have included a panel with \il{\delta_G=0}
whose features can be understood directly via the abrupt approximation.
Indeed, as expected, we see that all $\rhobar$ contours in the \il{\delta_G=0} case
are contours of either constant $\mbar_X t_G$, constant $\mbar_X/M_c$,
or constant $M_c t_G$.
The subsequent panels then illustrate how these contours deform as the width $\delta_G$
of our phase transition increases.

Likewise, in Fig.~\ref{fig:KKaxion3}, we plot the late-time tower fraction $\etabar$
of our KK axion tower.
Once again, we see a consistent picture emerging as a function of $\delta_G$. 
Indeed, just as for $\rhobar$ in Fig.~\ref{fig:KKaxion2}, we see that all $\etabar$ contours for \il{\delta_G=0} 
are contours of either constant $\mbar_X t_G$, constant $\mbar_X/M_c$,
or constant $M_c t_G$, 
with the subsequent panels once again illustrating how these contours deform 
as $\delta_G$ increases.
From the left panel of Fig.~\ref{fig:KKaxion3} we see that 
when the phase transition is abrupt, 
the total energy density of the tower
is distributed amongst a significant number of KK modes in
the small-$\Lambda_G$, small-$\hat f_X$ corner of parameter space. 
However, as our mass-generating phase transition unfolds over a longer period of time,
the energy contributions from the higher modes are suppressed and $\etabar$ begins to fall. 

One interesting feature apparent in Fig.~\ref{fig:KKaxion3}
is that the value of $\etabar$ is non-monotonic
as a function of $\Lambda_G$, first increasing and reaching a maximum near \il{\Lambda_G\approx 3}~GeV
before decreasing again. 
This can be understood as follows.
The vanishing of $\etabar$ for \il{\mbar_X \ll M_c} is understood from the distribution of initial abundances as described 
in Fig.~\ref{fig:initialabruptabundance}.
Indeed, in this region, nearly all of the abundance of the KK tower is contributed by the lowest KK mode because
very little mixing is generated in the phase transition.
By contrast, the decline in $\etabar$ which occurs as $\Lambda_G$ increases (or equivalently as $t_G$ decreases) 
can be understood as corresponding to our entrance 
into what in Refs.~\cite{DDM1,DDM2,DDMAxion}
was called the ``staggered'' regime wherein the heavier KK modes
          are oscillating (and thus already dissipating their energy density) 
as soon as our phase transition occurs, whereas the lighter KK modes remain overdamped for a long time after $t_G$.
This then results in  a total abundance composed primarily of contributions from those lighter modes.

Comparing the results in Figs.~\ref{fig:KKaxion2} and~\ref{fig:KKaxion3}, we see that the 
regions of parameter space which produce the largest total energy densities $\bar\rho$ almost 
coincide exactly with the regions that produce the largest tower fractions $\bar\eta$.   
Indeed, both regions have very small $\hat f_{X}$, and the only difference is that the 
former region has small $\Lambda_G$ while the latter region has \il{\Lambda_G\approx 3}~GeV.   
At first glance, this difference may appear to violate the claims made in connection with 
Fig.~\ref{fig:rhofracpiesgrand}, namely that these two regions should coincide completely.  However, 
the correlation observed in connection with Fig.~\ref{fig:rhofracpiesgrand} holds when $t_G$ 
is held fixed.   By contrast, the results in Figs.~\ref{fig:KKaxion2} and~\ref{fig:KKaxion3} 
are generated with $t_G$ implicitly varying throughout the parameter space shown.

Having focused in this section on the specific situation in which 
our $\Phi$ field is an axion, one possible next step would be to place
phenomenological bounds on the parameter spaces we have considered.
Such bounds would in principle mirror those which are normally applied
to the case of a traditional QCD axion, and come from a variety of considerations
including
supernova cooling,
black hole superradiance,
overclosure constraints,
dark-matter and dark-energy constraints,
traditional axion searches (such as light shining through walls), {\it etc}\/.
Within the context of the abrupt (\il{\delta_G=0}) limit, such bounds on KK axion towers
are discussed in detail in Ref.~\cite{DDMAxion}. 
Although the determination of such bounds for general $\delta_G$ is beyond the scope of this paper,  
the results we have obtained here concerning the late-time energy densities of these
axion systems should play a critical role in helping to determine exactly
where these bounds lie for general $\delta_G$, and the extent to which the suppressions and enhancements
we have observed translate into a loosening or strengthening of those bounds 
beyond traditional expectations.

\FloatBarrier
\section{Discussion and Conclusions\label{sec:Conclusions}}


In this paper we have investigated the effects of dynamical mass generation on 
the cosmological abundances of the Kaluza-Klein modes associated with a 
bulk scalar field.  In particular, we have examined the non-trivial case in which 
a phase transition localized on a brane leads to time-dependent masses and mixings among
these KK modes.  We have found that both the total energy density of the full KK tower
and the distribution of that energy density across the individual KK modes 
are extremely sensitive to the details of the phase transition.  As a result, 
within different regions of model parameter space, these abundances can be 
significantly enhanced or suppressed relative to standard expectations ---
sometimes by many orders of magnitude.  We have also derived a variety of approximate 
scaling behaviors and analytic expressions for the energy densities of the KK modes 
as functions of the relevant model parameters.  
In order to illustrate the potentially significant impact 
that these effects can have on the late-time abundances of our scalars within the context 
of a concrete model, we have also applied our general results to the case of a bulk axion/
axion-like field.
Finally, as a by-product of our analysis, we have also developed an alternate ``UV-based''
effective truncation of KK theories which is physically different from the more traditional
``IR-based'' truncation commonly employed in the literature, yet yields 
the same higher-dimensional theory as the truncation
is lifted.
 
Depending on the identity of the scalar field in question,
our results can have a variety of phenomenological implications.
For example, in the case in which 
our bulk scalar is an axion or axion-like field,
one can easily imagine a number of 
phenomenological consequences.  Note that
in general,
cosmological abundances in this paper
have been generated through a two-step process:
the assumption of a non-zero VEV for one or more modes 
in the KK tower
followed by a mass-generating phase transition.
As such, given the specific form of the initial conditions 
we have adopted in Eq.~(\ref{eq:initialconditions}),
the method of abundance generation we have studied in this paper
is tantamount to misalignment production.
However, misalignment production is not the only mechanism through which a population of 
axion-like particles can be produced in the early universe.  For example,
spontaneous breaking of the global $U(1)_X$ symmetry of Sect.~\ref{sec:AxionintheBulk} 
can lead to the formation of a network of cosmic strings and other topological 
defects~\cite{DavisAxionsProdCosmicString1}.  If this breaking occurs before cosmic 
inflation, these defects are simply inflated away.  By contrast, if the breaking of 
$U(1)_X$ occurs after inflation, these defects retain a non-negligible energy density 
until late times, and their decays can therefore generate a potentially significant 
contribution to the relic abundance of the corresponding axion-like particles.  Indeed, 
this topic has been studied in detail for the specific case of a QCD 
axion~\cite{DavisAxionsProdCosmicString2,
DabholkarCosmicStringAxionBounds,AxionProdStringsWalls,AxionProdDomainWalls}, but
in principle applies to other axion-like particles as well.
 
These considerations are important because 
the phase-space distribution of a 
population of axions or axion-like particles produced via the decays of 
topological defects is significantly different 
from that generated via misalignment 
production~\cite{ChangHagmannSikivie,AxionProdStringsWalls,AxionProdDomainWalls}.
Thus, in cases in which the misalignment contribution to the 
overall late-time axion abundance is suppressed by the time-dependent masses and mixings 
we have discussed here, 
the phase-space distribution
of relic axions can be significantly altered --- especially
if the contribution from topological defects ends up dominating the total axion abundance.
Moreover, the suppression of the
misalignment contribution to this overall axion abundance can also serve to weaken
the overclosure bound on the axion-decay constant, and thus could potentially enlarge the allowed 
region of parameter space for the QCD axion.   

Our results concerning the effects of a mass-generating phase transition can also have significant implications 
for other new-physics scenarios.  For example, our results may provide a way of mitigating 
the cosmological moduli problem which arises in supergravity~\cite{ModuliProblemSUGRA} and in 
string theory~\cite{BanksModuliProblem1,deCarlosModuliProblem,BanksModuliProblem2}.
Indeed, such theories generically predict large numbers of neutral scalar 
fields --- so-called moduli --- with flat potentials, long lifetimes, and 
large (and even Planck-scale) VEVs.  On the one hand, a number of 
phenomenological considerations imply that some mechanism must exist through which 
a potential is generated for these moduli, rendering them massive.  On the other 
hand, once these fields acquire masses, they can potentially overclose
the universe or precipitate an unacceptably late period of reheating.  A suppression
of the collective energy density of such moduli due to time-dependent mixing effects
of the sort we have discussed here
could potentially provide a way of addressing these issues.   

Our results also have implications within the context of the Dynamical
Dark Matter (DDM) framework~\cite{DDM1,DDM2}, an alternative framework for dark-matter
physics in which a potentially vast ensemble of unstable particles contribute to the 
present-day dark-matter abundance and in which phenomenological constraints on the dark sector
are satisfied through a balancing between constituent decay widths and cosmological abundances 
across the ensemble.  
Dark-matter ensembles with these properties emerge naturally in a variety of contexts~\cite{DDM1,DDM2,Chialva:2012rq,Dienes:2016kgc,Dienes:2016vei,Jaketoappear},
and detection strategies for such ensembles are discussed in 
Refs.~\cite{Dienes:2012yz,Dienes:2012cf,Dienes:2013xff,Dienes:2014via,Dienes:2014bka,Boddy:2016fds,Boddy:2016hbp}.
Indeed, it has been shown that the KK modes associated with a bulk axion-like
field which receives its abundance via misalignment production constitute a viable DDM
ensemble~\cite{DDM2,DDMAxion}, 
The results we have obtained here are thus directly applicable
to DDM ensembles of this sort and lead to a suppression of the overall relic abundance
of the ensemble for large $\delta_G$ relative to the relic abundance which arises for the abrupt case considered in 
Refs.~\cite{DDM2,DDMAxion}.  This suppression can potentially widen the phenomenologically allowed
parameter space of such DDM models.  Moreover, we also note from the results 
in Fig.~\ref{fig:rhofracpiesgrand}
that there exists a correlation 
between the overall magnitude of the late-time energy density of the KK tower
and the degree to which its distribution across the different KK modes is
particularly ``DDM-like'' (\ie, shared non-trivially across many different KK modes), with larger
total abundances tending to correlate with increased DDM-like behavior at late times.
Thus, we see that we can control the degree to which our ensemble of KK states
is truly DDM-like at late times simply by adjusting phenomenological parameters 
such as $\mbar$ and $\delta_G$ 
associated with our mass-generating phase transition. 

The suppression of late-time scalar abundances due to time-dependent masses and mixings 
also has potentially
important implications for a broad range of additional scenarios involving weakly-coupled scalar 
particles which receive a non-negligible contribution to their relic abundances from 
non-thermal mass-generating phase transitions of the sort we have discussed here.
The reason is that in such scenarios,   
a non-negligible population of these particles can also be produced
thermally from scattering processes involving SM particles in the radiation bath.  
This production mechanism is often referred to as ``freeze-in''~\cite{FreezeIn}, 
and it also gives rise to a population of particles whose phase-space distribution 
differs significantly from that of the population generated 
by non-thermal mass-generating phase transitions.
For the QCD axion, the freeze-in contribution to the total relic abundance   
is typically quite small~\cite{TurnerThermalAxion,MassoThermalAxion,GrafThermalAxion} 
compared to the contributions from other sources, such as misalignment production 
and cosmic-string and domain-wall decay.  By contrast, for other exotic scalars, the
freeze-in contribution can be significant.  Thus, in scenarios in which both misalignment
production and freeze-in production are na\" ively expected to generate comparable 
contributions to the overall abundance of a particular weakly-coupled scalar particle, a
suppression of the former contribution due to time-dependent mixing could 
both modify overclosure bounds and alter the expected 
phase-space distribution of the relic scalar population.

In this paper we have focused on the contribution to the total abundance
of a collection of KK scalar modes that arises due to a time-dependent
mass-generating phase transition.  In so doing, however, we have disregarded the effects 
of the quantum fluctuations that arise for these fields during the inflationary epoch.
In general, fluctuations in the long-wavelength modes of light fields --- and, in particular, those 
modes whose wavelengths exceed the Hubble length during that epoch --- behave like 
vacuum energy until after inflation ends.  Thus, the energy density in these 
fluctuations survives inflation and yields an additional contribution to the 
abundance of any field with mass \il{\lambda_k \lesssim H_I}, where $H_I$ is the value 
of the Hubble parameter during inflation.  This additional contribution depends 
sensitively on the parameters of the inflationary model (for a review, see, \eg, 
Ref.~\cite{InflationFluctuations}) and in particular on $H_I$ itself.  In 
Ref.~\cite{DDMAxion}, it was shown that for a KK tower of axion-like fields, the 
contribution to the relic abundance from vacuum misalignment dominates over this  
additional contribution from fluctuations during inflation for sufficiently small 
$H_I$.  However, we note that in general, this contribution exists and for 
higher-scale inflationary models may have a significant impact on the overall
relic abundance of the KK tower.  

In the course of our analysis in this paper, we also derived results for truncated
KK towers consisting of only $N$ modes, where \il{N<\infty}.
In so doing,
we presented the results of this finite-$N$ case merely as a stepping
stone on the way towards understanding the properties of the full KK tower that emerges in the 
\il{N\rightarrow\infty} limit.  However, the results of our finite-$N$
analysis are also interesting in their own right.  Indeed, the mass-squared 
matrix for a tower of $N$ modes is similar to the mass matrix which emerges 
in certain ``moose''~\cite{Moose} or ``quiver''~\cite{Quiver} gauge theories --- 
theories which also contain only a finite number of modes yet which yield 
a deconstructed extra spacetime dimension in the \il{N\rightarrow\infty} 
limit~\cite{Deconstruction}.  Thus, the results we have obtained here for such finite-$N$ 
theories should provide guidance as to how time-dependent mass-generating phase transitions
and time-dependent mixings will affect the late-time 
energy densities for the collections of scalars which emerge in such moose or quiver
theories.

It is also worth emphasizing that the results we have derived in
this paper reflect the particular KK-derived mixing 
structure of our model.  As a result, we did not observe certain phenomena which may in principle 
arise for more general systems of scalar fields in the presence of time-dependent 
mass-generating phase transitions and mixings.
For example, it has been shown~\cite{TwoTimescales} that in similar scalar systems with
more general mixing patterns, parametric resonances can arise which 
are extremely sensitive to the widths and mixings associated with mass-generating 
phase transitions and which 
``pump'' energy density into a particular field or fields during such phase transitions.  
Such systems may also exhibit so-called ``re-overdamping'' phenomena~\cite{TwoTimescales} in which the
total energy density of our system exhibits a time dependence which transcends those
normally associated with vacuum-dominated or matter-dominated cosmologies. 
However, in order for such novel effects to arise, the 
dynamically generated contribution $\overline{\mathcal{M}}^2_{\rm gen}$ to the 
mass-squared matrix at late times must satisfy the criterion   
\begin{equation}
  \det\overline{\mathcal{M}}_{\rm gen}^2 ~<~ 0~,
\end{equation}
and this criterion is not satisfied within the parameter space of our KK model.  
However, it is easy to contemplate early-universe
scenarios involving multiple scalars in which 
this condition is satisfied.
In such scenarios, parametric resonances and re-overdamping phenomena 
can then have a significant further impact on both the 
total abundance of the scalar fields involved and the distribution of that
total abundance among these scalars. 

There are a number of potentially interesting generalizations
and extensions of this work.  For example, in this paper we have focused on 
the case in which the higher-dimensional scalar propagates in a single, flat 
extra dimension.  However, such a flat extra dimension can be viewed as a
limit of the more general case of a warped geometry in which the 
extra dimension represents a slice of anti-de Sitter space~\cite{RS1,RS2}.  Such 
warped compactifications have a variety of phenomenological applications, such as
providing ways of addressing the hierarchy between 
the weak and Planck scales~\cite{RS1} as well as the hierarchies amongst the masses of 
the SM fermions~\cite{TonyRSYukawas1,HuberShafiRSYukawas}.
It would therefore be interesting to examine how our findings generalize to the 
case of a warped extra dimension.  It would also be interesting to 
consider the further generalization of our results to the case of multiple extra 
dimensions, both flat and warped.  
One could also imagine adopting more general initial conditions for our KK modes
beyond those in Eq.~(\ref{eq:initialconditions}), such as might arise if
there is an earlier phase of non-trivial dynamics (either in the bulk
or on the brane) prior to our mass-generating phase transition.

Another possible avenue for generalization concerns the nature of the cosmological
epoch during which we are presuming our dynamics to take place.
For example, throughout this paper we have focused primarily on the case in which the 
universe is effectively matter-dominated (\ie, with \il{\kappa = 2}) throughout the
period of mass generation.   Indeed, as discussed in Sect.~\ref{sec:TheModel},
phenomenological constraints on theories with large extra spacetime dimensions 
are more easily satisfied in LTR scenarios, in which the universe remains 
matter-dominated until very late times.  However, these constraints are 
considerably weaker in cases in which the compactification scale $M_c$ is 
$\mathcal{O}(\mathrm{TeV})$ or above.  Thus, it would also be interesting to 
consider the case in which the universe is effectively radiation-dominated 
(\ie, with \il{\kappa = 3/2}) during the mass-generation epoch.
Cosmologies with other values of $\kappa$ can be considered as well.

Finally, other interesting extensions of this work involve considering a broader
variety of cosmological contexts in which mass generation for
our KK modes might take place.
Throughout this paper, we have implicitly operated under the assumption that
the total energy 
density $\rho$ of the KK tower is negligible compared to the critical density 
$\rho_{\mathrm{crit}}$ during the mass-generation epoch.  Under such an assumption, the 
back-reaction of the KK tower on the evolution of the spacetime metric is therefore 
negligible during this epoch.  Thus, to a very good approximation, we may 
view the dynamics of the KK modes studied in this paper
as taking place within the context of a particular 
``background'' cosmology which is effectively decoupled from this dynamics and which
is therefore specified by a fixed functional form for the Hubble parameter $H(t)$ as
a function of $t$.  Indeed, such assumptions are valid for a variety of light scalars which
receive their abundances from misalignment production --- including, by necessity,
any such fields which serve as dark-matter candidates.  
However, one could alternatively consider the opposite regime in which 
\il{\rho\sim\rho_{\mathrm{crit}}} during the mass-generation epoch.  Under such conditions,
the back-reaction of the KK tower on $H(t)$ can be significant and must be 
incorporated into the equations of motion for the KK modes.  This situation 
can arise, for example, in scenarios in which these scalar particles play a role
in inflationary dynamics or in which they are responsible for additional, later 
periods of reheating after inflation in cosmologies with non-thermal
histories~\cite{WatsonNonThermalHistories}.  It would therefore be interesting
to study the evolution of the energy densities of the KK modes of our theory
in such scenarios.

\bigskip


\section*{Acknowledgments}


We would like to thank S.~Watson for useful discussions. 
The research activities of KRD and JK were supported in part by the Department
of Energy under Grant DE-FG02-13ER41976 (DE-SC0009913), while
the research activities of KRD were also supported in part by
the National Science Foundation through its employee IR/D program.
The opinions and conclusions
expressed herein are those of the authors, and do not represent any funding agencies.



\end{document}